\documentstyle[12pt]{article}
  
\catcode`@=11
\def\secteqno{\@addtoreset{equation}{section}%
\def\theequation{\thesection.\arabic{equation}}}
\catcode`@=12
\secteqno
\newcommand{\be}{\begin{equation}}
\newcommand{\ee}{\end{equation}}
\newcommand{\bea}{\begin{eqnarray}}
\newcommand{\eea}{\end{eqnarray}}
\newcommand{\bref}[1]{(\ref{#1})}
\newcommand{\nn}{\nonumber}

\newcommand{\C}[1]{{\cal #1}}
\newcommand{\mapright}[1]{\smash{[mathop{\hbox to 1cm{\rightarrowfill}}
\limits^{#1}}}

\newcommand{\T}{e^{-\phi}}
\newcommand{\N}{\mathcal{N}}
\newcommand{\eps}{\epsilon}
\newcommand{\del}{\partial}
\begin{document}

\begin{flushright}
\parbox{4.2cm}
{2012,~May 29\\
KEK-TH-1529 \hfill \\
}
\end{flushright}

\vspace*{1.1cm}

\begin{center}
 \Large\bf Canonical approach to Courant brackets for D-branes
\end{center}
\vspace*{1.5cm}
\centerline{\large Machiko Hatsuda$^{\ast\dagger a}$ and Tetsuji Kimura$^{\ast b}$}
\begin{center}
$^{\ast}$\emph{KEK Theory Center, High Energy Accelerator Research 
Organization,\\
Tsukuba, Ibaraki 305-0801, Japan} 
\vspace*{0.5cm}
\\
$^{\dagger}$\emph{Physics Department, Juntendo University, 270-1695, Japan}
\vspace*{1cm}
\\
$^{a}$mhatsuda@post.kek.jp
~~;~~
 $^{b}$tetsuji@post.kek.jp
\end{center}

\vspace*{1cm}

\centerline{\bf Abstract}
  
\vspace*{0.5cm}
We present an extension of the Courant bracket to the ones for D$p$-branes
by analyzing Hamiltonians and local superalgebras.
Contrast to the basis of the bracket for a fundamental string 
which consists of the momentum and  the winding modes,
the ones for D$p$-branes contain higher rank R-R coupling tensors.  
We show that the R-R gauge transformation rules
 are obtained by these Courant brackets for D$p$-branes
where the  Dirac-Born-Infeld gauge field and the
``two-vierbein field" play an essential role. 
Canonical  analysis of the worldvolume theories 
naturally gives the basis of the brackets and  the target space 
backgrounds keeping T-duality manifest  at least  for NS-NS sector.
In a D3-brane analysis S-duality is manifest 
as a symmetry of interchanging 
the NS-NS coupling and the R-R coupling.  
\vfill 

\thispagestyle{empty}
\setcounter{page}{0}
\newpage 
\section{Introduction and summary}

In string theories the general coordinate trasformation symmetry is enlarged to
the gauge symmetry for the Kalb-Ramond field 
in addition to the gravitational field.
This is an inevitable feature from T-duality of string theories
which mixes the gravitational field and the Kalb-Ramond field.
T-duality has a long history of the research 
\cite{Buscher:1987sk,Duff:1989tf,Maharana:1992my}.
Along the study of T-duality as a target space duality
Siegel wrote down the gauge transformation rule of 
the gravitational field and the Kalb-Ramond field,
$G_{mn}$ and $B_{mn}$, in a T-duality covariant way
\cite{Siegel:1993xq,Siegel:1993th,{Siegel:1993bj}}.
 Hitchin introduced the generalized geometry
 with the Courant bracket 
which gives this gauge transformation involving $B_{mn}$ field
 \cite{Hitchin:2004ut}.
Hull introduced the doubled  formalism
with manifest T-duality to flux compactifications
by introducing non-geometry
 \cite{Hull:2004in}.
Hull and Zwiebach used the closed string field theory
to construct the double field theory
defined by the C-bracket which is reduced to the Courant bracket
\cite{Hull:2009mi}.

In bosonic string theory
both T-duality and the gauge symmetry of $G_{mn}$ and $B_{mn}$
 are governed by O(d,d) symmetry consistently.
The momentum and the winding modes of a string are
 the building block of O(d,d) vector.
On the other hand in type II superstring theory there are R-R gauge fields.
They couple to D-branes whose charges are transformed 
as a spinor under SO(d,d). 
T-duality interchanges IIA D-branes
and IIB D-branes. Furthermore
the IIB theory  includes S-duality.
Then T-duality is enlarged to ``U-duality"
\cite{HT,Obers:1998fb}.
Corresponding to this enlargement 
the gauge symmetry involving R-R gauge fields
should be enlarged.
M theory is a powerful theory to explore 
U-duality and 
the enlarged gauge symmetry  cooperative to U-duality
\cite{Hull:2007zu,Berman:2011cg}.
There is also an approach from the supergravity 
theory \cite{{Pacheco},Coimbra:2011nw}.
In this paper we focus on a D-brane extension of the
gauge transformation given by a new type of Courant bracket,
leaving the U-duality problem for D-branes.
We clarify the background field  dependence of 
Hamiltonians for D-branes, and reveal 
differences between the fundamental string case
and D-brane cases.

We take a canonical approach of worldvolume theories
to explore the enlarged gauge symmetry for D-branes.
For the fundamental string
the momentum and the winding modes
construct an O(d,d) vector  $Z_M=(p_m,~\partial_\sigma x^m)$.
The canonical Hamiltonian ${\cal H}_\perp$ 
and the $\sigma$-diffeomorphism constraint 
${\cal H}_\parallel$  are 
expressed in terms of the basis 
as ${\cal H}_\perp=Z_M{\cal M}^{MN}Z_N$
and ${\cal H}_\parallel=Z_M\tilde{\eta}^{MN}Z_N$
with an off-diagonal O(d,d) invariant metric
 $\tilde{\eta}^{MN}$.
The target space background fields are included only in ${\cal M}^{MN}$.
This expression has manifest T-duality symmetry.
On the other hand  the gauge symmetry 
is generated by $Z_M$.
The canonical  bracket between $Z_M$'s  makes 
a closed algebra including a stringy anomalous term.
This anomalous term is proportional to the
O(d,d) invariant metric.
The Courant bracket is obtained by reading off from the regular 
coefficient of this canonical algebra 
\cite{Siegel:1993xq,Siegel:1993th,Alekseev:2004np}.
The gauge transformation of 
$G_{mn}$ and $B_{mn}$
is given by the Courant bracket between
the ``two-vierbein field" and a gauge parameter,
where the    ``two-vierbein field" is 
an O(d,d) vector representation of  $G_{mn}$ and $B_{mn}$
\cite{Siegel:1993xq}.
This is an ideal O(d,d) vector for both
T-duality and the gauge symmetry.

Now the problem is an extension of this analysis to D$p$-branes.  
There are proposals of extensions of the Courant bracket
for $p$-branes in \cite{Hull:2007zu,{Pacheco}}.
When we try to extend it in canonical approach
we face to two crucial differences  from string:
1. A  D$p$-brane has $p$ worldvolume spatial directions,
so replacing  $\partial_\sigma x^m$ 
by $\partial_i x^m$ with $i=1,\cdots, p $ 
does not work out straightforwardly.
2. R-R gauge fields are higher rank tensors whose treatment in  
the framework of the Courant bracket is unknown. 
These problems are partially resolved by the help of the Dirac-Born-Infeld 
(DBI) U(1) gauge field and the two-vierbein field.
For the first problem the cotangent vector corresponding to 
the winding mode is constructed as $E^i\partial_i x^m$
where $E^i$ is DBI gauge field strength \cite{Hatsuda:1997pq,{Kamimura:1997ju},{Hatsuda:1998by}}. 
For the second problem we found that
the basis of the Courant bracket for D$p$-brane 
consists of the higher rank tensors.
The R-R gauge fields build a vector in this enlarged space
by contracting with the two-vierbein field.
Then we show that the R-R gauge transformation rules 
are generated by our Courant brackets for D$p$-branes
which contains  Chern-Simons terms.
In the reference \cite{Pacheco} the exceptional Courant bracket
contains Chern-Simons terms. 
In our approach the Chern-Simons terms are obtained from the canonical commutator
between DBI gauge fields.

This paper is organized as follows:
In section 2 we analyze the gauge generator algebra 
and the Courant bracket for 
a fundamental (F) string, and we extend it to the one for  a D3-brane
from their local superalgebras in flat space.
Several extensions of the Courant bracket 
to the one involving $p$-form were introduced in \cite{Bonelli:2005ti}.
In section 3 background fields are taken into account in this formulation 
for a F-string.
We also show that a string on a group manifold such as  AdS space
has the same structure as the  Courant bracket.
The similar bracket was introduced in 
\cite{Siegel:1993bj,Grana:2012rr}.
For a D-string we demonstrate how the R-R coupling
causes differences from the F-string case, and 
we present an extension of the Courant bracket to reproduce
the gauge transformation of the R-R gauge field.
In section 4 the above analysis is extended for IIA D$p$-branes and IIB D$p$-branes.
Obtained basis of  Courant brackets and 
background matrices are subsets of whole U-duality.   
In order to construct a U-duality manifest theory 
these subsets will be combined in some sense.

There are many interesting researches on this subject;
generalized geometry
 to flux compactifications in physics \cite{Grana:2005jc},
 recent work on double field theory 
\cite{Hohm:2010pp} and
doubled formalism and D-branes 
\cite{Albertsson:2008gq,Albertsson:2011ux}.

\par\vskip 6mm
\vskip 6mm
\section{Flat background }

 D-brane is solitonic excitation in type II superstring theories
and the BPS condition is given by the N=2 supersymmetry.
The BPS mass is determined from the global supersymmetry algebra,
while local information such as  the Virasoro
condition is determined from the local superalgebra.
The local superalgebra is written in terms of supercovariant derivatives,
 $(d_{A\alpha},~p_m\pm \partial_\sigma x^m+\cdots)$ in flat space 
with spacetime vector index $m$, spinor index $\alpha$, and
N=2 supersymmetry index $A$.
We can read off a complete set of the bosonic basis of the R-R coupling
from the right hand side of the local superalgebra, 
$\{d_{A\alpha},d_{B\beta}\}$, even for a flat space case.
In this section we begin by local superalgebras in flat space 
to extract the basis to describe the Hamiltonian and the $\sigma$
diffeomorphism constraint, which are the Virasoro constraints.
From the canonical bracket of these basis
we construct a Courant brackets.
Then we show the gauge symmetry transformation rules for
R-R gauge fields by using this Courant bracket.

\subsection{F-string  }

The local superalgebra for a IIB F-string in
flat space is given as
\bea
\{d_{A\alpha}(\sigma) ,d_{B\beta}(\sigma')\}
&=&
\left[\delta_{AB}{}p_{\alpha\beta}(\sigma)
+(\tau_3)_{{AB}}
\partial_\sigma x{}_{\alpha\beta}(\sigma)\right]
\delta(\sigma-\sigma')
\nn\\
&=&Z_M(\sigma)\Gamma^M_{AB;\alpha\beta}\delta(\sigma-\sigma')~~~,
\label{localsusy}
\eea
with $p_{\alpha\beta}=p_m \gamma^m{}_{\alpha\beta}$, 
$x_{\alpha\beta}=x^m \gamma_m{}_{\alpha\beta}$ and  
$Z_M=(p_m,~\partial_\sigma x^m)$.
$(\Gamma^M)_{AB;\alpha\beta}=
\left(\delta_{AB}\gamma^m{}_{\alpha\beta},~
\tau_3{}_{AB}\gamma_{m}{}_{\alpha\beta}
\right)$ 
is a gamma matrix for the type II 
theories satisfying $\{\Gamma^M,\Gamma^N\}=2\tilde{\delta}^{MN}$.
In the right hand side of \bref{localsusy}
fermionic coordinates  are set to be zero,
and we will focus only on the bosonic part 
in this paper. 
The canonical bracket is given by   
$\{p_m(\sigma),\partial_\sigma x^n(\sigma')\}=i\delta^n_m
\partial_\sigma \delta(\sigma-\sigma')$,
where a curly bracket $\{*,*\}$ is used  for a
distinction from the Courant bracket during this paper.

Hamiltonian is a linear combination of the
$\tau$-diffeomorphism ${\cal H}_\perp$ 
and the $\sigma$ diffeomorphism constraints ${\cal H}_\parallel$.
For simplicity we call ${\cal H}_\perp$  Hamiltonian
from now on, where it is the case in the conformal gauge.    
The Hamiltonian and the $\sigma$-diffeomorphism constraint are given by 
\bea
\left\{
\begin{array}{cclcc}
{\cal H}_\perp &=&\displaystyle\frac{1}{2}
\frac{1}{32}{\rm tr}(Z_M \Gamma^M)^2&&\\
&=&\displaystyle\frac{1}{2}Z_M\tilde{\delta}^{MN}Z_N~=~
\displaystyle\frac{1}{2}\left(
{p}{}^2+(\partial_\sigma x){}^2
\right)~=~0&,&\tilde{\delta}^{MN}=
\left(\begin{array}{cc}
\eta^{mn}&0\\0&\eta_{mn}
\end{array}\right)\\
{\cal H}_\parallel&=&\displaystyle\frac{1}{2}Z_M\tilde{\eta}^{MN}Z_N
~=~ \partial_\sigma x{}^m{p}_m~=~0&,&
\tilde{\eta}^{MN}=
\left(\begin{array}{cc}
0&\delta_m^n\\\delta_n^m&0
\end{array}\right)
\end{array}\right.
\label{HamSt}
\eea
where $\tilde{\eta}^{MN}$ is the O(d,d) invariant metric while 
$\eta^{mn}$ and $\eta_{mn}$ are d-dimensional Minkowski metrics.

Let us consider a geometry generated by $Z_M$ which satisfies the following algebra
\bea
\left\{Z_M(\sigma),Z_N(\sigma')\right\}=i\eta_{MN}\partial_\sigma \delta(\sigma-\sigma')
~~,~~{\eta}_{MN}=
\left(\begin{array}{cc}
0&\delta_m^n\\\delta_n^m&0
\end{array}\right)~~~.
\label{ZZdel}
\eea  
The right hand side is the stringy anomalous term, which
is proportional to the O(d,d) invariant metric $\eta_{MN}$. 
$\tilde{\eta}^{MN}$ and $\eta_{MN}$ are 
introduced independently, however they coincide with each other
for the F-string case  relating the gauge symmetry and the T-duality consistently.
The ``operator" $Z_M$ is supposed to act as the derivative
with respect to ``double field space coordinates" \cite{{Hull:2009mi}}
as
\bea
\left\{Z_M,f(x,\tilde{x})\right\}_{double}=-i\partial_M f
\equiv \left(\partial_m f,~ \partial^m f\right)
=-i
\left(\displaystyle\frac{\partial}{\partial x^m}f,~
\displaystyle\frac{\partial}{\partial \tilde{x}_m}f
\right)~~~.\label{opZ}
\eea
Local parameters $\Lambda(x)$'s  are introduced in a vector form
$\hat{V}=V^MZ_M$ as
\bea
\hat{\Lambda}(\sigma)=\Lambda^M Z_M
=\Lambda^m p_m+\Lambda_m \partial_\sigma x^m=(\Lambda^m,~\Lambda_m)
~~~.\label{lambdasigma}
\eea
The canonical commutator between two $\hat{\Lambda}$'s is calculated as
\cite{Alekseev:2004np}
\bea
\{\hat{\Lambda}_1(\sigma),\hat{\Lambda}_2(\sigma')\}
\label{lam1lam2}
&=&
-i\left(\Lambda_{[1}^M\partial_M\Lambda_{2]}^NZ_N
-\frac{1}{2}\Lambda_{[1}{}^M\partial_\sigma \Lambda_{2]M}
-{K}\partial_\sigma \Psi_{(12)}
\right)\delta(\sigma-\sigma')\\
&&
+i\left(
(\frac{1}{2}+K)\Psi_{(12)}(\sigma)
+(\frac{1}{2}-K)\Psi_{(12)}(\sigma')
\right)
\partial_\sigma
\delta(\sigma-\sigma')\nn
\eea
with a symmetric product
\bea
\Psi_{(12)}=\Lambda_{(1}^m\Lambda_{2)m}=\frac{1}{2}\Lambda_{(1}^M\Lambda_{2)}^N\eta_{NM}=
\frac{1}{2}\Lambda_{(1}^M\Lambda_{2)M}~~~.
\eea 
The O(d,d) invariant metric $\eta_{MN}$ is used for lowering indices $_M$.
The coefficient $K$ is an arbitrary number 
corresponding to an ambiguity of the total derivative
$\partial_\sigma \Psi_{(12)}$ 
caused from a term containing $\partial_\sigma \delta(\sigma-\sigma')$. 
The condition 
  $\partial^m \Lambda=0 $
  leads to $\partial_\sigma \Lambda=
\partial_\sigma x^m\partial_m \Lambda=
Z_M\eta^{MN}\partial_N \Lambda$.
Then the commutator \bref{lam1lam2} becomes the $Z_M$ algebra 
with anomalous term
\bea
\{\hat{\Lambda}_1(\sigma),\hat{\Lambda}_2(\sigma')\}&=&
-i\hat{\Lambda}_{12}(\sigma)\delta(\sigma-\sigma')\nn\\
&&+i\left(
(\frac{1}{2}+K)\Psi_{(12)}(\sigma)
+(\frac{1}{2}-K)\Psi_{(12)}(\sigma')
\right)
\partial_\sigma
\delta(\sigma-\sigma')\nn
\nn\\
\hat{\Lambda}_{12}&=&\Lambda_{12}^NZ_N\nn\\
\Lambda_{12}^N&=&
\Lambda_{[1}^M\partial_M\Lambda_{2]}{}^N
-\frac{1}{2}\Lambda_{[1}{}^M\partial^N \Lambda_{2]M}
-{K}\partial^N\Psi_{(12)}~~~.
\label{L12}
\eea
The regular coefficient of 
\bref{L12}  is called  the C-bracket in   a 
double field space, $\Lambda_{12}{}^N=\left(
[\Lambda_1,\Lambda_2]_C\right)^N$.
Especially  
\bea
\left(
[\hat{\Lambda}_1,\hat{\Lambda}_2]_C\right)^N
~=~\left\{
\begin{array}{lcl}
\Lambda_{[1}^M\partial_M\Lambda_{2]}{}^N
-\displaystyle\frac{1}{2}\Lambda_{[1}{}^M\partial^N \Lambda_{2]M}
&\cdots&K=0\\\\
\Lambda_{1}^M\partial_M\Lambda_{2}{}^N
+\Lambda_{2}{}^M\partial_{[L|} \Lambda_{1|M]}\eta^{LN}
&\cdots&K=-\displaystyle\frac{1}{2}
\end{array}
\right.
\eea
The Jacobiator  of the algebra is calculated 
in terms of the doubled indices as 
\bea
&\left[\left[\displaystyle\int\hat{\Lambda}_1, \displaystyle\int\hat{\Lambda}_2\right]_C,
 \displaystyle\int\hat{\Lambda}_3
\right]_C+{\rm cyclic~sum}
=
-\displaystyle\int\hat{\Lambda}_{[123]}&\nn\\
&\hat{\Lambda}_{[123]}=
{\Lambda}_{[123]}{}^NZ_N=\displaystyle\frac{1}{4}\partial^N\left(
{\Lambda}_{[1}^L {\Lambda}_{2}^M\partial_M {\Lambda}_{3]}{}_L\right)Z_N
=\partial_\sigma\left(\displaystyle\frac{1}{4}
{\Lambda}_{[1}^L {\Lambda}_{2}^M\partial_M {\Lambda}_{3]}{}_L\right)
&~~~,\label{Jacobiator}
\eea
which is independent of the value $K$.
The breakdown of the Jacobi identity is given by a
total derivative term
 so it does not cause serious inconsistency in general.

The gauge symmetry generator is invariant 
under a further gauge symmetry
\bea
\delta \Lambda^M&=&\partial^M \zeta \nn\\
\delta \displaystyle\int \hat{\Lambda}&=&
\displaystyle\int
 Z_M \partial^M \zeta =
i\int d\sigma \{{\cal H}_\parallel, \zeta \}_{double}=
i \int d\sigma ~ \partial_\sigma \zeta~~~, \label{dsigma}
\eea
which vanishes for a closed string. 
In the second equality of \bref{dsigma}
we used the fact that
 ${\cal H}_\parallel $ is the $\sigma$-diffeomorphism constraint
in \bref{HamSt} and  \bref{opZ}.
When the parameter satisfies $\partial^m \zeta =0$,
the double field space bracket is reduced to the usual canonical bracket.

The C-bracket is reduced  to the Courant bracket under the assumption 
 $\partial^m \lambda=0$.
Let us introduce a tangent vector $\lambda\in T $ and 
a cotangent vector $\lambda^\ast\in T^\ast $ 
\bea
&&
\hat{\Lambda}=
\lambda+\lambda^\ast~~,~~
\lambda=\Lambda^m p_m~~,~~
\lambda^\ast=\Lambda_m d x^m\nn\\
&&\left[
\hat{\Lambda}_1,\hat{\Lambda}_2
\right]_{Courant}
~=~[\lambda_1,\lambda_2]+{\cal L}_{\lambda_1}\lambda_2^\ast-
{\cal L}_{\lambda_2}\lambda_1^\ast
-\displaystyle\frac{1}{2}d(\iota_{\lambda_1} \lambda_2^\ast-\iota_{\lambda_2} \lambda_1^\ast)
\label{F1flatLambda}
\\
&&~~
\left\{\begin{array}{ccl}
[\lambda_1,\lambda_2]&=&\Lambda_{[1}^m\partial_m\Lambda_{2]}^n p_n\\
{\cal L}_{\lambda_{1}}\lambda_{2}^\ast&=&
\left(\Lambda_{1}^m\partial_m\Lambda_{2;n} +
(\partial_n\Lambda_{1}^m)\Lambda_{2;m}
\right)\partial_\sigma x^n\\
d(\iota_{\lambda_{1}}\lambda^\ast_{2})&=&
\partial_n
(\Lambda_{1}{}^m \Lambda_{2;m})
\partial_\sigma x^n
\end{array}\right.~~~\nn
\eea
for $K=0$. 
The Courant bracket for $K=-1/2$ is given by;
\bea
&&\left[\hat{\Lambda}_1,\hat{\Lambda}_2\right]_{Courant}
~=~[\lambda_1,\lambda_2]
+{\cal L}_{\lambda_1}\lambda_2^\ast
-\iota_{\lambda_2}d\lambda_1^\ast
\label{F1flatLambdahalf}\\
&&~~~~~~~\left\{~\iota_{\lambda_2}d\lambda_1^\ast~=~
\Lambda_{2}{}^m\partial_{[n|} \Lambda_{1|m]} ~ \partial_\sigma x^n
~~~.\nn\right.
\eea
The last term is the gauge transformation of the antisymmetric gauge field.

It was shown in \cite{Siegel:1993xq,Siegel:1993th} that 
the gauge transformation rule of $G_{mn}$ and $B_{mn}$
are given by the bracket 
between the ``two-vierbein vector" $(e_a{}^m,~e_{ma})$ 
and a gauge parameter vector
$(\xi^m,~\xi_m)$. 
Originally Siegel called  the bracket a ``new Lie derivative"
which is recognized as the C-bracket \cite{Hull:2009mi}. 
  The two-vierbein fields are transformed linearly under O(d,d)
 leading to
the  fractional linearly  transformation of $G_{mn}+B_{mn}$ 
\cite{Duff:1989tf,Maharana:1992my} as:
\bea
&&G_{mn}+B_{mn}=e_{mn}=e_{ma}e_n{}^a
~,~e_a{}^m e_m{}^b=\delta_a^b,~
~ e_m{}^ae_a{}^n=\delta_m^n~~\label{GB}\\\nn\\
&&\hat{e}_a
~\to~
 \hat{e}'_a=
  \left(\begin{array}{c}e'_a{}^m{}\\e'_{ma}\end{array} \right)
 = 
 \left(\begin{array}{cc}
 A^m{}_l&B^{ml}\\
C_{ml}&D_m{}^l 
 \end{array} \right) 
\left(\begin{array}{c}e_a{}^l\\e_{la}\end{array} \right),~
\left(\begin{array}{cc}A&B\\C&D\end{array}\right)\in
{\rm O(d,d)}
\nn\\\nn\\
&&\left(\begin{array}{c}e'_a{}^m\\e'_{ma}\end{array} \right)
e'_n{}^a
=\left(\begin{array}{c}
\delta_n^m\\e_{mn}'\end{array} \right)
~
\Rightarrow~~e'_{mn}=\left(C_{ml}+D_m{}^{p}e_{pl}\right)\left(A^n{}_l+B^{nq}e_{ql}\right)^{-1}
 ~~~.
\eea
 The Courant bracket  \bref{F1flatLambdahalf}
 bewteen the two-vierbein field and the gauge parameter
is given by
\bea
&&\delta_\xi \hat{e}_a~=~\left[
\hat{\xi},\hat{e}_a\right]_{Courant}~~,~~
\hat{e}_a=\left(\begin{array}{c}e_a{}^m\\e_{ma}\end{array}\right)~,~
\hat{\xi}=\left(
\begin{array}{c}\xi^m\\\xi_m\end{array}\right)
\nn\\
&&~\left\{{\renewcommand{\arraystretch}{1.2}
\begin{array}{ccl}
\delta_\xi e_a{}^m&=&\xi^n\partial_n e_a{}^m-e_a{}^n\partial_n\xi^m\\
\delta_\xi e_{ma}&=&\xi^n\partial_n e_{ma}
+e_{na}\partial_m \xi^n+
e_a{}^n\partial_{[m}\xi_{n]}
\end{array}}\right.\label{2vier}~~~.
\eea
From the relation \bref{GB} the transformation \bref{2vier} gives 
the gauge transformation rules for $G_{mn}$ and $B_{mn}$  as
\bea
\Rightarrow
&&~\left\{
{\renewcommand{\arraystretch}{1.2}
\begin{array}{ccl}
\delta_\xi G_{mn}&=&\xi^l\partial_l G_{mn}+
\partial_{(m|}\xi^lG_{l|n)}\\
\delta_\xi B_{mn}&=&\xi^l\partial_l B_{mn}
+\partial_{[m|}\xi^lB_{l|n]}+\partial_{[m}\xi_{n]}
\end{array}}\right.~~.\label{dGB}
\eea
This two-vierbein formalism is essential to extend D-brane systems
as we will see in the next subsection.

\subsection{D3-brane }

The local superalgebra for D3-brane in flat space is given by
\cite{Kamimura:1997ju}
\bea
&&\{
d_{A\alpha}(\sigma) ,d_{B\beta}(\sigma')\}
\nn\\
&&~~=~
\left[\delta_{AB}{}p_{\alpha\beta}(\sigma)
+(\tau_3)_{{AB}}
q^{NS}_{\alpha\beta}(\sigma)
+(\tau_1)_{AB}q^{R;[1]}_{\alpha\beta}(\sigma)
+(i\tau_2)_{{AB}}q^{R;[3]}_{\alpha\beta}(\sigma)
+c_{AB\alpha\beta}\Phi (\sigma)\right]
\delta(\sigma-\sigma')
\nn\\
&&~~=~Z_M(\sigma)\Gamma^M_{AB;\alpha\beta}\delta(\sigma-\sigma')~~~,\\
&&~~~~~~~~{\renewcommand{\arraystretch}{1.2}
\begin{array}{lcl}
p_{\alpha\beta}=p_m(\gamma^m)_{\alpha\beta}&,&
q^{R;[1]}_{\alpha\beta}=\epsilon^{ijk} 
F_{ij}\partial_kx{}^m(\gamma_m){}_{\alpha\beta}\\
q^{NS}_{\alpha\beta}
= E^\i\partial_ix{}^m (\gamma_m){}_{\alpha\beta}
&,&
q^{R;[3]}_{\alpha\beta}= \epsilon^{ijk}
\partial_ix^m \partial_j x^n \partial_k x^l 
(\gamma_{mnl}){}_{\alpha\beta}
\end{array}}
~,~
\nn
\eea
with the Gauss law constraint 
\bea
\Phi&=&\partial_i E^i=0~~~.
\eea
$E^i$ and $F_{ij}=\partial_{[i}A_{j]}$ 
are Dirac-Born-Infeld 
(DBI) U(1) electric field and magnetic field
on a D3-brane and $c_{AB\alpha\beta}$ is a function.
We focus on the bosonic part only.
Let us consider the algebra generated by
\bea
&Z_M=\left(
\begin{array}{c}
p_m\\
E^i \partial_i x^m\\\hline
\epsilon^{ijk}F_{ij}\partial_kx^m
\\
\epsilon^{ijk}\partial_ix^m 
\partial_jx^n 
\partial_kx^l
\end{array}
\right)&~~~\label{ZD3}
\eea
which is a vector of $T\oplus T^\ast \oplus 
\Lambda^1 T^\ast \oplus \Lambda^3T^\ast$ \cite{Hull:2007zu}.
The upper half is the NS-NS sector and  lower half is the R-R sector.
The normalization in \bref{ZD3} is omitted, 
while the correct coefficient 
is given by \bref{coefficient1} and \bref{coefficient2} in the appendix.
The Hamiltonian is the sum of bilinears in $Z_M$
\bea
\left\{\begin{array}{ccl}
{\cal H}_\perp&=&\displaystyle\frac{1}{2}~
\frac{1}{32}{\rm tr}(Z_M \Gamma^M){}^2
~=~\frac{1}{2}Z_M\tilde{\delta}^{MN}Z_N
\\\\
&=&\displaystyle\frac{1}{2}
\left(
{p}^2+(E^i \partial_i x)^2+
(\epsilon^{ijk}F_{ij}\partial_kx^m)^2+
(\epsilon^{ijk}\partial_ix^m\partial_jx^n\partial_kx^l)^2
\right)=0\\ \\
{\cal H}_i &=&
 \partial_ix^m p_m+F_{ij} E^j=0
\end{array}\right.\nn~~~.
\eea
The worldvolume diffeomorphism constraints
 ${\cal H}_i=0$ can be written 
in a bilinear form of $Z_M$  
by contracting with $E^i$, $\epsilon^{ijk}F_{jk}$ and 
 $\epsilon^{ijk}\partial_jx^m\partial_kx^n$ as
\bea
&&E^i {\cal H}_i=
\epsilon^{ijk}F_{ij} {\cal H}_k=
\epsilon^{ijk}\partial_i x^n \partial_j x^l 
{\cal H}_k=0\nn\\
&&\Rightarrow
~~Z_M\tilde{\rho}^{MN}Z_N=0~~,~~
\tilde{\rho}^{MN}
=\left(
\begin{array}{cc|cc}
0&a\delta_n^m&b\delta_n^m&c_{[n_1n_2}\delta_{n]}^m\\
a\delta_m^n&0 &c_{mn}&0\\\hline
b\delta_m^n&c_{mn}&&\\
c_{[m_1m_2}\delta_{m]}^n
&0&&
\end{array}
\right)
\label{rhotildeD3}
\eea
with arbitrary coefficients $a,b,c_{mn}$.

The algebra generated by $Z_M$ is closed 
by the  Gauss law constraint.
The D3-brane extension of \bref{ZZdel} is given by 
\bea
&&
\left\{Z_M(\sigma),Z_N(\sigma')\right\}
=i{\rho}_{MN}^i \partial_i \delta^{(3)}(\sigma-\sigma')\nn\\
\nn\\
&&{\rho}_{MN}^i=
\left(
{\renewcommand{\arraystretch}{1.2}
\begin{array}{cc|cc}
0& E^i \delta_m^n&\epsilon^{ijk}F_{jk}\delta_m^n&
\frac{1}{2} \epsilon^{ijk}\partial_{j} x^{[n_1}
\partial_k x^{n_2}\delta_m^{n]}\\
E^i \delta_n^m&0&\epsilon^{ijk}\partial_j x^{[m}\partial_kx^{n]}&0\\\hline
\partial^{ijk}F_{jk}\delta_n^m&
\epsilon^{ijk}\partial_j x^{[n}\partial_kx^{m]}&&\\
\frac{1}{2}\epsilon^{ijk}\partial_j
 x^{[m_1}\partial_k x^{m_2}\delta_n^{m]}&0&&
\end{array}
}
\right)\nn\\
\label{ZZrho}
\eea
with ${\cal O}_{[abc]}={\cal O}_{a[bc]}+
{\cal O}_{b[ca]}+{\cal O}_{c[ab]}$.
The Gauss law and the Bianchi identity guarantee 
that the right hand side of \bref{ZZrho} is a total derivative.  
Unlike the F-string case,
for a D-brane case  $\tilde{\rho}^{MN}$ in \bref{rhotildeD3}
 relating to T-duality
and $\rho_{MN}^i$ in \bref{ZZrho} relating to gauge symmetry neither coincide
nor manifest the O(d,d) symmetry.

Let us write down a canonical commutator between two vectors 
\bea
\hat{\Lambda}_I(\sigma)~=~ \Lambda_I{}^MZ_M~=~
(\Lambda_{I}{}^n,~\Lambda_{I}{}_n,~\tilde{\Lambda}_{I}{}_n,~ \Lambda_{I}{}_{nrs})
\eea
as
\bea
&&\left\{\hat{\Lambda}_1(\sigma),\hat{\Lambda}_2(\sigma')\right\}
~=~
-i\hat{\Lambda}_{12}(\sigma)\delta^{(3)}(\sigma-\sigma')\nn\\
&&~~~~~~~~~~~~~~~~~~~~~~~~~~~+i\left((\frac{1}{2}+K)\Psi_{(12)}^i(\sigma)
+(\frac{1}{2}-K)\Psi_{(12)}^i(\sigma')
\right)
\partial_i\delta^{(3)}(\sigma-\sigma')\label{LD3}
\\
&&\left\{
{\renewcommand{\arraystretch}{1.2}
\begin{array}{ccl}
\Lambda_{12}{}^n&=&\Lambda_{[1}{}^m\partial_m \Lambda_{2]}{}^n\nn\\
\Lambda_{12;}{}_n&=&\Lambda_{[1}{}^m\partial_m \Lambda_{2]}{}_n
-\frac{1}{2}(\Lambda_{[1}^m\partial_n\Lambda_{2]m}
+\Lambda_{[1|m}\partial_n\Lambda_{|2]}^m)
-K\partial_n\left(\Lambda_{(1}^m\Lambda_{2)m}\right)\nn\\
\tilde{\Lambda}_{12;}{}_n&=&\Lambda_{[1}{}^m\partial_m \tilde{\Lambda}_{2]}{}_n
-\frac{1}{2}(\Lambda_{[1}^m\partial_n\tilde{\Lambda}_{2]m}
+\tilde{\Lambda}_{[1|m}\partial_n\Lambda_{|2]}^m)
-K\partial_n\left(\Lambda_{(1}^m\tilde{\Lambda}_{2)m}\right)\nn\\
 \Lambda_{12;}{}_{nlr} &=&
 \Lambda_{[1}{}^m\partial_m \Lambda_{2]nlr}
-\frac{1}{4}\Bigl(\Lambda_{[1}^m\partial_{[n|} \Lambda_{2]m|lr]}
-(\partial_{[n|} \Lambda_{[1}^m ) \Lambda_{2]m|lr]} \Bigr)
-\frac{K}{2}
\partial_{[n|}
(\Lambda_{(1}^m \Lambda_{2)m|lr]} )\nn\\
&&+\frac{1}{6}\left(
\Lambda_{[n}^{[1}\partial_{l}\tilde{\Lambda}_{r]}^{2]}
+\partial_{[n}\Lambda_{l}^{[1}\tilde{\Lambda}_{r]}^{2]}
\right)
-\frac{K}{3}\partial_{[n}(\Lambda^{(1}_{l}\tilde{\Lambda}_{r]}^{2)})\nn\\
\Psi_{(12)}^i&=&\frac{1}{2}\Lambda^M_{(1}\Lambda^N_{2)}\rho_{MN}^i
\end{array}}\right.
   \label{L12Cbra}~~~.
\eea
where $\Psi_{(12)}^i$ is an ambiguity caused from 
$\partial_i\delta^{(3)}(\sigma-\sigma')$.

Now we refer to the coefficient $\hat{\Lambda}_{12}(\sigma)$ in \bref{LD3} as
 the Courant bracket for  D3-brane 
analogously to the previous section.
A vector in $Z_M$ space is denoted by
\bea
& 
\hat{\Lambda}=
\lambda+\lambda^\ast+\lambda^{[1]}+\lambda^{[3]}
~\in~
T \oplus T^\ast \oplus \Lambda^{1}T^\ast  \oplus \Lambda^{3}T^\ast 
&\nn
\eea
then  $\hat{\Lambda}_{12}(\sigma)$ in \bref{LD3} 
is  recognized as 
the Courant bracket for  D3-brane  with $K=0$ as ;
\bea
[\hat{\Lambda}_1,\hat{\Lambda}_2]_{D3}
&=&[\lambda_1,\lambda_2]+{\cal L}_{\lambda_{[1}}\lambda_{2]}^\ast
+{\cal L}_{\lambda_{[1}}\lambda_{2]}^{[1]}
+{\cal L}_{\lambda_{[1}}\lambda_{2]}^{[3]}
-\frac{1}{2}d
\left(
\iota_{\lambda_{[1}}\lambda^\ast_{2]}
+\iota_{\lambda_{[1}}\lambda^{[1]}_{2]}
+\iota_{\lambda_{[1}}\lambda^{[3]}_{2]}
\right)
\nn\\
&&+\frac{1}{6}\left(
\lambda_{[1}^\ast \wedge d\lambda_{2]}^{[1]}
+d\lambda_{[1}^{\ast}\wedge \lambda_{2]}^{[1]}
\right)\label{0CS}
\eea
\bea
&\left\{
{\renewcommand{\arraystretch}{1.2}
\begin{array}{ccl}
[\lambda_1,\lambda_2]&=&\Lambda_{[1}^m\partial_m\Lambda_{2]}^n~p_n\\
{\cal L}_{\lambda_{1}}\lambda_{2}^\ast&=&\left(\Lambda_{1}^m\partial_m\Lambda_{2;n} +
(\partial_n\Lambda_{1}^m)\Lambda_{2;m}
\right)~E^i\partial_i x^n\\
{\cal L}_{\lambda_{1}}\lambda_{2}^{[1]}&=&
\left(\Lambda_{1}^m\partial_m\tilde{\Lambda}_{2;n} +
(\partial_n\Lambda_{1}^m)\tilde{\Lambda}_{2;m}
\right)~
\epsilon^{ijk}
F_{ij}\partial_k x^n\\
{\cal L}_{\lambda_{1}}\lambda_{2}^{[3]}&=&
\left(\Lambda_{1}^m\partial_m\Lambda_{2;nlr} +
\frac{1}{2}(\partial_{[n|}\Lambda_{1}^m)\Lambda_{2;m|lr]}
\right)~
\epsilon^{ijk}\partial_ix^n
\partial_jx^l
\partial_kx^r\\
d(\iota_{\lambda_1}\lambda^\ast_2)&=&
\partial_n(
\Lambda_1^m\Lambda_{2;m})~E^i\partial_i x^n\\
d(\iota_{\lambda_1}\lambda_2^{[1]})&=&
\partial_n(
\Lambda_1^m\tilde{\Lambda}_{2;m})~
\epsilon^{ijk}F_{ij} \partial_kx^n\\
d(\iota_{\lambda_1}\lambda_{2}^{[3]})&=&
\partial_n(\Lambda_1^m\Lambda_{2;mlr})~
\epsilon^{ijk}\partial_i 
x^n\partial_j x^l \partial_k x^r
\\
\lambda_{1}^\ast\wedge d\lambda_{2}^{[1]}&=&
\Lambda^1_{[n}\partial_{l}\tilde{\Lambda}^2_{r]}
\epsilon^{ijk}\partial_i x^n\partial_j x^l
\partial_k x^r\\
d\lambda_{1}^\ast\wedge \lambda_{2}^{[1]}&=&
(\partial_{[n}\Lambda^1_{l})\tilde{\Lambda}^2_{r]}
\epsilon^{ijk}\partial_i x^n\partial_j x^l
\partial_k x^r\\
\end{array}}\right.&
\eea
It is also convenient to introduce
the Courant bracket for  D3-brane with $K=-1/2$;
\bea
[\hat{\Lambda}_1,\hat{\Lambda}_2]_{D3}
&=&[\lambda_1,\lambda_2]
+{\cal L}_{\lambda_{1}}\lambda_{2}^\ast
+{\cal L}_{\lambda_{1}}\lambda_{2}^{[1]}
+{\cal L}_{\lambda_{1}}\lambda_{2}^{[3]}
-
\iota_{\lambda_{2}}d\lambda^\ast_{1}
-\iota_{\lambda_{2}}d\lambda^{[1]}_{1}
-\iota_{\lambda_{2}}d\lambda^{[3]}_{1}
\nn
\\
&&+\frac{1}{3}\left(
\lambda_{2}^{[1]}\wedge d\lambda_1^\ast
-\lambda_2^\ast\wedge d\lambda_1^{[1]}
\right)\label{12CS}
\eea
\bea
&\left\{
{\renewcommand{\arraystretch}{1.2}
\begin{array}{ccl}
\iota_{\lambda_{2}}d\lambda^\ast_{1}&=&
\Lambda_2^m\partial_{[m|}\Lambda_{1;|n]}~E^i\partial_ix^n
\\
\iota_{\lambda_{2}}d\lambda^{[1]}_{1}&=&
\Lambda_2^m\partial_{[m|}\tilde{\Lambda}_{1;|n]}
~\epsilon^{ijk}F_{ij}\partial_kx^n\\
\iota_{\lambda_{2}}d\lambda^{[3]}_{1}&=&
\frac{1}{3!}\Lambda_2^m\partial_{[m|}{\Lambda}_{1;|nlr]}
\epsilon^{ijk}\partial_i x^n
\partial_j x^l \partial_kx^r
\end{array}}\right.&\label{CourantD3}
\eea
In our canonical approach the appearence of the Chern-Simons terms,  
as shown to exist in \cite{Pacheco},
  in the second lines of 
\bref{0CS} and \bref{12CS} 
comes from the canonical commutator between the DBI U(1) fields.

The Jacobi identity is also broken by  a
total derivative term,
since the D3 extended space vector $\hat{\Lambda}$ 
can be also written as 
\bea
 \hat{\Lambda}=\Lambda^m p_m
+{\Lambda}_m^i \partial_i x^m 
~~,~~
{\Lambda}_m^i=
\Lambda_m E^i +\tilde{\Lambda}_m \epsilon^{ijk}F_{jk}+
\Lambda_{mnl}\epsilon^{ijk}\partial_j x^n \partial_k x^l
\eea
which leads to a total derivative term as the Jacobiator 
as seen in \bref{Jacobiator}.

Next we examine the gauge transformation rules for the 
R-R gauge fields.
We extend the NS-NS gauge fields to the R-R gauge fields for D3-brane as
\bea
\hat{e}_a=\left(
\begin{array}{c}
e_a{}^m\\e_{ma}
\end{array}\right)~~\Rightarrow~~
\hat{C}_a^{D3}=\left(
{\renewcommand{\arraystretch}{1.2}
\begin{array}{c}
e_a{}^m\\e_{ma}=e_a{}^l(B_{ml}+G_{ml})\\\hline
e_a{}^l C_{ml}^{[2]}\\
e_a{}^p C_{mnlp}^{[4]}
\end{array}}\right)~~~.
\eea
Then the gauge transformations 
for the R-R gauge fields 
are given by the Courant bracket in \bref{CourantD3}
with the parameter as
\bea
&\delta_\xi\hat{C}^{D3}_a=\left[\hat{\xi},\hat{C}^{D3}_a\right]_{D3}~~,~~
\hat{\xi}=\left(
\begin{array}{c}
\xi^m\\
\xi_m\\\hline
{\xi}_m^{[1]}
\\
{\xi}_{mnl}^{[3]}
\end{array}\right)&
\nn\\
&\Rightarrow~~~
\left\{
{\renewcommand{\arraystretch}{1.2}
\begin{array}{ccl}
\delta_\xi C^{[2]}&=&{\cal L}_\xi C^{[2]}+d\xi^{[1]}\\
\delta_\xi C^{[4]}&=&{\cal L}_\xi C^{[4]}-
C^{[2]}\wedge d\xi+d\xi^{[3]}+B\wedge d\xi^{[1]}
\end{array}}\right.&
\label{gaugeRR}
\eea
where $\xi=\xi_m$ is the gauge parameter for $B$ field.
The gauge transformation of $C^{[4]}$ involves the $B$ field
as expected.
\par \vskip 6mm

\vskip 6mm
\section{Strings in curved background}

In the previous section we have shown three points:
the canonical approach to the Hamiltonian; the local superalgebra
gives a Courant bracket for D3-brane;
 and the gauge symmetry
of the R-R gauge fields are obtained by our Courant bracket.
The two-vierbein formalism of 
the NS-NS gauge fields, $G_{mn}$ and $B_{mn}$,
constructs the  O(d,d) vector
manifesting the T-duality transformation and the gauge transformation.
A similar mechanism seems to work for the R-R gauge fields.
In order to analyze the R-R couplings  directly 
we begin by strings in a curved background
clarifying the background fields dependence.

\subsection{F-string }

The bosonic part of the action for a F-string 
in a curved space is given by
\bea
I&=&\int d^2\sigma~{\cal L}~~,~~
{\cal L}~=~{\cal L}_{NG}+{\cal L}_{WZ}\nn\\
{\cal L}_{NG}&=&-T_{F1}\sqrt{-h}~~,~~
h=\det h_{\mu\nu}~~,~~h_{\mu\nu}=\partial_\mu x^m \partial_\nu x^n G_{mn} \nn\\
{\cal L}_{WZ}&=&\frac{1}{2}T_{F1}\epsilon^{\mu\nu}\partial_\mu x^m \partial_\nu x^n
B_{mn} ~~~.\nn
\eea
where $T_{F1}=\frac{1}{2\pi\alpha'}$ and $B_{mn}$ is the NS-NS
two-form gauge field.
The canonical momentum is defined as
\bea
p_m&\equiv&\frac{\partial {\cal L}}{\partial (\partial_0 x^m)}
=-T_{F1}\sqrt{-h}h^{0\mu}\partial_{\mu}x^n
G_{mn}+T_{F1}\epsilon^{01}\partial_\sigma x^n B_{mn}~~~,
\eea
where $h^{\mu\nu}$ is the inverse of $h_{\mu\nu}$.

The Hamiltonian is   obtained by the Legendre transformation as
\bea
&&H~=~\int d\sigma ~{\cal H}\nn\\
&&{\cal H}~=~p_m\partial_{0}x^m-{\cal L}
~=~-\frac{1}{\sqrt{-h}h^{00}}{\cal H}_\perp
-\frac{h^{01}}{h^{00}}{\cal H}_\parallel \nn\\
&&\left\{
{\renewcommand{\arraystretch}{1.2}
\begin{array}{ccl}
{\cal H}_\perp &=&\displaystyle\frac{1}{2T_{F1}}\left(
\tilde{p}_m G^{mn}\tilde{p}_n
+T_{F1}{}^2 h_{11}
\right)~=~0\nn\\
{\cal H}_\parallel&=& \partial_\sigma x{}^m  \tilde{p}_m~=~
\partial_\sigma x{}^m  {p}_m~=~
0
\end{array}}\right.
\nn
\eea
with 
\bea
\tilde{p}_m&\equiv&p_m-T_{F1}\epsilon^{01}\partial_\sigma x^n B_{mn}~~~.
\eea
By virtue of the  bilinear expression $ h_{11}=x'{}^mG_{mn}x'{}^n$,
${\cal H}_\perp$ is recast into the sum of bilinears as
\bea
{\cal H}_\perp&=&\frac{1}{2T_{F1}}
\left(\tilde{p}_m~~~T_{F1}\partial_\sigma x^m{}\right)
\left(
\begin{array}{cc}
G^{mn}&0\\
0&G_{mn}
\end{array}
\right)
\left(
\begin{array}{c}
\tilde{p}_n\\T_{F1} \partial_\sigma x^n{}
\end{array}
\right)\nn\\
&=&
\frac{1}{2T_{F1}}
Z_M{}^T~{\cal M}^{MN}~Z_N\nn\\
&&\left\{\begin{array}{ccl}
Z_N&=&\left(
\begin{array}{c}
p_n\\T_{F1} \partial_\sigma x^n
\end{array}
\right)\\ \\
{\cal M}^{MN}&=&
\left(
\begin{array}{cc}
\delta^m{}_p&0\\B_{mp}&\delta_m{}^p
\end{array}
\right)
\left(
\begin{array}{cc}
G^{pq}&0\\0&G_{pq}
\end{array}
\right)
\left(
\begin{array}{cc}
\delta_q{}^n&-B_{qn}\\0&\delta^q{}_n
\end{array}
\right)\\\\
&=&\left(
\begin{array}{cc}
G^{mn}&-G^{mq}B_{qn}\\
B_{mp}G^{pn}&G_{mn}-
B_{mp}G^{pq}B_{qn}
\end{array}
\right)
\end{array}\right.~~~.\label{F1T}
\eea
The $Z_M=Z_M(\sigma)$ base contains only the worldvolume variables,
while ${\cal M}^{MN}={\cal M}^{MN}(x)$ 
contains only the spacetime background fields.
It is further written as
\bea
\begin{array}{c}
{\cal M}^{MN}=\tilde{\delta}^{AB}E_A{}^ME_B{}^N\\
G^{pq}=\eta^{ab}e_a{}^pe_b{}^q\\
G_{pq}=\eta_{ab}e^a{}_pe^b{}_q
\end{array}
~~,~~
E_{A}{}^M
=\left(
\begin{array}{cc}
e_a{}^q&0\\
0&e^a{}_q
\end{array}
\right)
\left(
\begin{array}{cc}
\delta_q{}^n&-B_{qn}\\
0&\delta^q{}_n
\end{array}
\right)~~~,
\eea
and 
\bea
{\cal H}_\parallel=\frac{1}{2}Z_M\tilde{\eta}^{MN}Z_N~~,~~
\tilde{\eta}^{MN}=\left(
\begin{array}{cc}
0&\delta^m_n\\
\delta_m^n&0
\end{array}
\right)~~~.
\eea
There exists the O(d,d) symmetry which preserves $\tilde{\eta}^{MN}$
for a string background.
The T-duality transformation of the background fields for a bosonic string 
is the O(d,d) transformation.
The subgroup which preserve both $\tilde{\eta}^{MN}$ and $\tilde{\delta}^{MN}$
is O(d)$\times$O(d).
 $E_A{}^M$  is a coset element of O(d,d)/O(d)$\times$O(d) whose dimension is
the same as GL(d) as well as the one of  
the two-vierbein field $\hat{e}_a$.

\vskip 6mm
\subsection{AdS-string}

It is possible to extend  the subsection 2.1 to a string
on  AdS (or sphere) space which is described by a group manifold
G.
For a group element $X\in $G 
 the left invariant one-form and the covariant derivative 
are given as 
\bea
&J^a=X^{-1}dX= dx^m e_m{}^a~~,~~
D_a=e_a{}^m p_m~~,~~e_m{}^a e_a{}^n=\delta_m^n~~,~~
~ e_a{}^me_m{}^b=\delta_a^b~~.
&
\eea 
The Hamiltonian and the $\sigma$-diffeomorphism constraint
are written in terms of
$Z_A=(~D_{ a},J^{a}~)$ as
\bea
&&\left\{
{\renewcommand{\arraystretch}{1.2}
\begin{array}{ccl}
{\cal H}_\perp &=&\frac{1}{2}Z_A \tilde{\delta}^{AB}Z_B
=\displaystyle\frac{1}{2}\left(
(D_a){}^2+(J^a)^2\right)~=~0\nn\\
{\cal H}_\parallel&=&\frac{1}{2}Z_A\tilde{\eta}^{AB}Z_B=
J{}^a{D}_a = \partial_\sigma x^m  p_m=0
\end{array}}\right.
\eea
with $\tilde{\delta}^{AB}$=diag($\eta^{ab},~\eta_{ab}$).

Let us consider a space generated by 
$Z_A$
satisfying the following algebra \cite{Hatsuda:2001xf}
\bea
&&\{Z_A(\sigma), Z_B(\sigma')\}=-iF_{AB}{}^CZ_C\delta(\sigma-\sigma')
+i\eta_{AB}\partial_\sigma\delta(\sigma-\sigma')
\label{AdSZZ}\\
&&~~~F_{AB}{}^CZ_C=\left(\begin{array}{cc}
f_{ab}{}^cD_c&-f_{ac}{}^b J^c\\
f_{bc}{}^aJ^c&0
\end{array}\right)~~,~~
\eta_{AB}=\left(\begin{array}{cc}
0&\delta_a^b\\
\delta_b^a&0
\end{array}\right)~~~\nn\\
&&~~~f_{ab}{}^c=e_{[a}{}^m\partial_m e_{b]}{}^n e_n{}^c\nn
\eea
where $f_{ab}{}^c$ is the structure constant for the group G.
A vector $\hat{\Lambda}$ in the space  $T\oplus T^* $
    is introduced 
\bea
\hat{\Lambda}(\sigma)=\Lambda^A Z_A=
\Lambda^a D_a+\Lambda_a J^a =
\Lambda^a e_a{}^m p_m+ \Lambda_a e_m{}^a \partial_\sigma x^m
~~,
\eea
then the canonical bracket between two vectors is given by
\bea
\{\hat{\Lambda}_1(\sigma), \hat{\Lambda}_2(\sigma') \}&=&
-i\hat{\Lambda}_{12}\delta(\sigma-\sigma')+
i\left((\frac{1}{2}+K)\Psi_{(12)}(\sigma)+
 (\frac{1}{2}-K)\Psi_{(12)}(\sigma')\right)
\partial_\sigma \delta(\sigma-\sigma')
\nn\\
{\Lambda}_{12}{}^A&=&\frac{1}{2}\Lambda_{[1}^C\Lambda_{2]}^BF_{CB}{}^A
+\Lambda_{[1}{}^B \partial_B \Lambda_{2]}^A-\frac{1}{2}
\Lambda_{[1}^B \partial^A \Lambda_{2]B} 
-K\partial^A\Psi_{12}
\nn\\
\Psi_{(12)}&=&\Lambda_{(1}^a\Lambda_{2)a}=\frac{1}{2}\Lambda_{(1}^M\Lambda_{2)M}\nn
\eea
with $\{Z_A,\Lambda\}=-i\partial_A \Lambda=-i
(e_a{}^m\partial_m \Lambda, e_m{}^a\partial^m \Lambda )$.
Now the C-bracket in the double field space is $
(\left[\hat{\Lambda}_1,\hat{\Lambda}_2
\right]_C)^A
={\Lambda}_{12}{}^A$
which contains the structure constant.
A similar bracket is introduced in \cite{Siegel:1993bj,Grana:2012rr}.
If we denote
$(\Lambda^a D_a,~\Lambda_a J^a) =
(\Lambda^a e_a{}^m p_m,~ \Lambda_a e_m{}^a \partial_\sigma x^m)
=(\lambda,~\lambda^\ast)$ and 
impose $\partial^m \Lambda=0=\partial^a \Lambda$, then the  C-bracket
is reduced  to the Courant bracket given by \bref{F1flatLambda} as
\bea
&&\left[\hat{\Lambda}_1,~\hat{\Lambda}_2 \right]_{AdS}=
[\lambda_1,\lambda_2]+{\cal L}_{\lambda_{[1}}\lambda^\ast_{2]}
-\frac{1}{2}d(\iota_{\lambda_{[1}}\lambda^\ast_{2]})\nn\\
&&~~~
\left\{
{\renewcommand{\arraystretch}{1.2}
\begin{array}{ccl}
[\lambda_1,\lambda_2]&=&
\left(\Lambda_{[1}^a\partial_a\Lambda_{2]}^b+
\frac{1}{2}\Lambda^a_{[1}\Lambda_{2]}^cf_{ac}{}^b\right)D_b\\
{\cal L}_{\lambda_{[1}}\lambda_{2]}^\ast&=&
\left(\Lambda_{[1}^a\partial_a\Lambda_{2]b}
-\Lambda^a_{[1}\Lambda_{2]c} f_{ab}{}^c
+\partial_b\Lambda_{[1}^a\Lambda_{2]a}
\right)J^b\\
d(\iota_{\lambda_{1}}\lambda^\ast_{2})&=&
\partial_a(\Lambda_{1}^b\Lambda_{2;b})J^a
\end{array}}
\right.~~~.
\eea
 \vskip 6mm
\vskip 6mm
\subsection{D-string}

The bosonic part of the action for a D-string 
in a curved space is given by \footnote{There is an alternative formulation for a D-brane \cite{Bakhmatov:2011ab}.}
\bea
I&=&\int d^2\sigma~{\cal L}~~,~~{\cal L}={\cal L}_{DBI}+{\cal L}_{WZ}
\label{D1act}\\
{\cal L}_{DBI}&=&-T_{D1}e^{-\phi}\sqrt{-h_{F}}~~,~~
h_{F}=\det h_F{}_{\mu\nu}\nn\\
{\cal L}_{WZ}&=&\frac{T_{D1}}{2}\epsilon^{\mu\nu}
\left(
\partial_\mu x^m \partial_\nu x^n C_{mn}^{[2]}
+2\pi\alpha'{\cal F}_{\mu\nu} C^{[0]}
\right)
\nn\\
h_F{}_{\mu\nu}&=&\partial_\mu x^m \partial_\nu x^n (G_{mn}+
B_{mn})+2\pi \alpha' F_{\mu\nu} \nn\\
F_{\mu\nu}&=&\partial_\mu A_\nu-\partial_\nu A_\mu ~~,~~
{\cal F}_{\mu\nu}~=~F_{\mu\nu}+\frac{1}{2\pi\alpha'}
\partial_\mu x^m\partial_\nu x^n B_{mn}
~~~.\nn
\eea
The canonical momenta are defined as 
\bea
p_m&\equiv&\displaystyle\frac{\partial {\cal L}}{\partial (\partial_0 x^m)}
\nn\\
&=&-T_{D1}e^{-\phi}\sqrt{-h_F}\left(\frac{1}{2}h_F{}^{(\mu 0)}G_{mn}+
\frac{1}{2}h_F{}^{[\mu 0]}
B_{mn}\right)\partial_\mu x^n
+T_{D1}\epsilon^{01}\partial_\sigma x^n
\left( C_{mn}^{[2]}+C^{[0]}B_{mn}
\right)\nn\\
E^1&\equiv&\displaystyle\frac{\partial {\cal L}}{\partial (\partial_0 A_1)}
\nn\\
&=&2\pi\alpha' T_{D1}\left(
-e^{-\phi}\sqrt{-h_F}\frac{1}{2}h_F{}^{[10]}+\epsilon^{01}C^{[0]}\right)~~~.
\eea
A matrix $h_F{}^{\mu\nu}$ is the inverse of $h_F{}_{\mu\nu}$
as $h_F{}_{\mu\nu}h_F{}^{\nu\rho}=\delta_\mu^{\rho}$.
Symmetrized/antisymmetrized indices are denoted as 
$h_F{}^{(\mu 0)}=h_F{}^{\mu 0}+h_F{}^{0\mu}$
and $h_F{}^{[\mu 0]}=h_F{}^{\mu 0}-h_F{}^{0\mu}$.
The Legendre transformation brings the Lagrangian to the 
Hamiltonian as
\bea
H&=&\displaystyle\int d\sigma ~{\cal H}\nn\\
{\cal H}&=&p_m\partial_0 x^m+E^1\partial_0 A_1-{\cal L}\nn\\
&=&-\frac{1}{\sqrt{-h}h^{00}}{\cal H}_\perp
-\frac{h^{01}}{h^{00}}{\cal H}_\parallel 
-A_0~\Phi\nn\\
&&\left\{\begin{array}{ccl}
{\cal H}_\perp &=&\displaystyle\frac{1}{2T_{D1}}e^\phi\left(
\tilde{p}_m G^{mn}\tilde{p}_n
+\frac{1}{(2\pi\alpha')^2}
\tilde{E}^1h_{11}\tilde{E}^1
+T_{D1}{}^2e^{-2\phi}  h_{11}
\right)~=~0\nn\\
{\cal H}_\parallel&=&\tilde{p}_m\partial_\sigma  x^m~=~
{p}_m\partial_\sigma  x^m~=~
0\\
\Phi&=&\partial_\sigma E^1~=~0
\end{array}\right.
\nn
\eea
with 
\bea
\tilde{p}_m&\equiv&p_m
-B_{mn}\frac{1}{2\pi\alpha'}E^1\partial_\sigma x^n
-C_{mn}^{[2]}T_{D1}\epsilon^{01}\partial_\sigma x^n \nn\\
&=&-T_{D1}e^{-\phi}
\sqrt{-h_F}\frac{1}{2}h_F{}^{(\mu 0)}G_{mn}\partial_\mu x^n~~\nn\\
\tilde{E}^1&\equiv&E^1-C^{[0]}2\pi\alpha' T_{D1}\epsilon^{01}=-\frac{2\pi\alpha'}{2} T_{D1}e^{-\phi}\sqrt{-h_F}h_F{}^{[10]}~.
\eea

Analogously to the previous section
${\cal H}_\perp$ is recast into the sum of bilinears as
\bea
&&{\cal H}_\perp~=~\frac{1}{2T_{D1}}
Z_M{}^T~{\cal M}^{MN}~Z_N\nn
\eea
with 
\bea
Z_N&=&\left(
{\renewcommand{\arraystretch}{1.4}
\begin{array}{c}
p_n\\
\frac{1}{2\pi \alpha'}E^1\partial_\sigma x^n{}
\\\hline 
T_{D1} \partial_\sigma x^n{}
\end{array}}
\right)\label{ZD1}
\eea
\bea
&&{\cal M}^{MN}\nn\\
&&\nn =
{\renewcommand{\arraystretch}{1.4}
\left(
\begin{array}{cc|c}
\delta^m{}_p&0&\\
B_{mp}&\delta_m{}^p&\\\hline
\tilde{C}^{[2]}_{mp}
&-C^{[0]}\delta_m{}^p
&\delta_m{}^p
\end{array}
\right)
\left(
\begin{array}{cc|c}
e^{\phi}G^{pq}&0&\\
0&e^\phi G_{pq}&\\\hline
&&e^{-\phi}G_{pq}
\end{array}
\right)
\left(
\begin{array}{cc|c}
\delta_q{}^n&-B_{qn}&-\tilde{C}_{qn}^{[2]}\\
0&\delta^q{}_n&-C^{[0]}\delta^q{}_n\\\hline
&&\delta^q{}_n
\end{array}
\right)}\\ \nn\\\nn
&&={\renewcommand{\arraystretch}{1.4}
\left(
\begin{array}{cc|c}
e^{\phi}G^{mn}&-e^{\phi}G^{mq}B_{qn}&-e^{\phi}G^{mq}\tilde{C}_{qn}^{[2]}\\
e^{\phi}B_{mp}G^{pn}
&e^{\phi}G_{mn}-e^{\phi}B_{mp}G^{pq}B_{qn}
&
-e^{\phi}C^{[0]}G_{mn}-e^{\phi}B_{mp}G^{pq}\tilde{C}^{[2]}_{qn}
\\
\hline
e^{\phi}\tilde{C}^{[2]}_{mp}G^{pn}
&
-e^{\phi}C^{[0]}G_{mn}-e^{\phi}\tilde{C}^{[2]}_{mp}G^{pq}B_{qn}
&\left(e^{-\phi}+e^{\phi}(C^{[0]})^2\right)G_{mn}
-e^{\phi}\tilde{C}_{mp}^{[2]}G^{pq}\tilde{C}_{qn}^{[2]}
\end{array}
\right)}\nn\\
\label{MD1}
\eea
with $\tilde{C}^{[2]}_{mn}=C^{[2]}_{mn}+C^{[0]}B_{mn}$.
The R-R coupling is separated as the above. It
has the inverse dilaton dependence $e^{-\phi}$.
The upper-left part of the ${\cal M}$ matrix is the same as the 
F-string case in \bref{F1T}.

The $Z_M$ algebra is given as
\bea
\left\{Z_M(\sigma),Z_N(\sigma')\right\}=i\rho_{MN}\partial_\sigma \delta(\sigma-\sigma')
~~,~~
\rho_{MN}=\left({\renewcommand{\arraystretch}{1.4}
\begin{array}{cc|c}
0&
\frac{1}{2\pi\alpha'}E^1\delta_m^n&T_{D1}\delta_m^n\\
\frac{1}{2\pi\alpha'}E^1\delta_n^m&0&0\\\hline
T_{D1}\delta_n^m&0&0
\end{array}}
\right)~~~.\nn\\
\eea
The canonical bracket between two vectors
$\hat{\Lambda}_I=(\Lambda_I^m,~\Lambda_{I;m},~
\tilde{\Lambda}_{I;m}
)~\in$~$ T\oplus T^\ast\oplus \Lambda^1 T^\ast$
is
\bea
\left\{\hat{\Lambda}_1(\sigma),\hat{\Lambda}_2(\sigma')\right\}&=&
-i\hat{\Lambda}_{12}\delta(\sigma-\sigma')
+i\left((\frac{1}{2}+K)\Psi_{(12)}(\sigma)+(\frac{1}{2}-K)\Psi_{(12)}(\sigma')
\right)\partial_\sigma\delta(\sigma-\sigma')\nn\\
\hat{\Lambda}_{12}&=&\Lambda_{[1}^m\partial_m\Lambda_{2]}^n~p_n\\
&&+\left(
\Lambda_{[1}^m\partial_m\tilde{\tilde{\Lambda}}_{2]n}
-\frac{1}{2}(\Lambda_{[1}^m\partial_n\tilde{\tilde{\Lambda}}_{2]m}
-\partial_n\Lambda_{[1}^m\tilde{\tilde{\Lambda}}_{2]m}
)-K\partial_n\Psi_{(12)}
\right)\partial_\sigma x^n\nn\\
\tilde{\tilde{\Lambda}}_{I;m}&=&\frac{E^1}{2\pi\alpha'}\Lambda_{I;m}+T_{D1}\tilde{\Lambda}_{I;m}\nn\\
\Psi_{(12)}&=&\Lambda_{(1}^m\tilde{\tilde{\Lambda}}_{2)m}~~~.\nn
\eea
Introducing the notation of vectors as
\bea
&\hat{\Lambda}_I=\lambda_I+\lambda_I^\ast+\lambda_I^{[1]}~\in~
T\oplus T^\ast\oplus \Lambda^1 T^\ast &,
\eea
we refer to the coefficient $\hat{\Lambda}_{12}$ as the 
Courant bracket for D-string 
\bea
&\left[\hat{\Lambda}_1,\hat{\Lambda}_2\right]_{D1}
=\left\{
{\renewcommand{\arraystretch}{1.4}
\begin{array}{lcc}
\left[\lambda_1,\lambda_2\right]+{\cal L}_{\lambda_{[1}}\lambda_{2]}^\ast
+{\cal L}_{\lambda_{[1}}\lambda_{2]}^{[1]}
-\displaystyle\frac{1}{2}d(\iota_{\lambda_{[1}}\lambda_{2]}^\ast+
\iota_{\lambda_{[1}}\lambda_{2]}^{[1]})&\cdots&K=0\\
\left[\lambda_1,\lambda_2\right]
+{\cal L}_{\lambda_{1}}\lambda_{2}^\ast
+{\cal L}_{\lambda_{1}}\lambda_{2}^{[1]}
-\iota_{\lambda_{2}}d\lambda_{1}^\ast-
\iota_{\lambda_{1}}d\lambda_{1}^{[1]}&\cdots&K=-\displaystyle\frac{1}{2}
\end{array}}
\right.&\nn\\
\eea
The Jacobiator follows from \bref{Jacobiator}
by replacing $\hat{\Lambda}_I=(\Lambda_I^m,~\tilde{\tilde{\Lambda}}_{I;m})$.

The gauge transformation of the R-R gauge field is
given by the Courant bracket between a 
gauge field vector $\hat{C}^{D1}$
and a parameter vector $\hat{\xi}$
\bea
&\hat{C}^{D1}_a=
{\renewcommand{\arraystretch}{1.2}
\left(
\begin{array}{c}e_a{}^m\\e_{ma}\\\hline e_a{}^n C_{mn}^{[2]}
\end{array}\right)
~~,~~
\hat{\xi}=\left(
\begin{array}{c}\xi^m\\\xi_m\\\hline{\xi}_m^{[1]}
\end{array}\right)}&\nn\\
&\delta_\xi\hat{C}^{D1}_a=\left[
\hat{\xi},\hat{C}^{D1}_a
\right]_{D1}
~\Rightarrow~
\delta_\xi C^{[2]}~=~{\cal L}_\xi C^{[2]}+d\xi^{[1]}
\label{D1}~~~
\eea
 as well as the one for the NS-NS gauge fields in \bref{dGB} .

\par\vskip 6mm
\vskip 6mm
\section{D$p$-branes in curved background}

In this section we analyze
 the Courant bracket 
for arbitrary type II D$p$-branes
including the background gauge field dependence.
The R-R gauge transformation rules are given by
the Courant brackets between parameter vectors and 
 gauge field vectors.
There are several studies for D$p$-brane on a doubled compact space
\cite{Albertsson:2008gq} and on a doubled 
non-compact space \cite{Albertsson:2011ux}.

\subsection{D2-brane}

The action for a D2-brane is an extension of \bref{D1act}
\bea
I&=&I_{DBI}+I_{WZ} ~~,~~I_{DBI}=\displaystyle\int_M d^2\sigma~{\cal L}_{DBI}
\nn\\
{\cal L}_{DBI}&=&-T_{D2}e^{-\phi}\sqrt{-h_{F}}~~,~~
h_{F}=\det h_F{}_{\mu\nu}\nn\\
{\cal L}_{WZ}&=&
 \displaystyle\frac{1}{3!}T_{D2}\epsilon^{\mu\nu\rho}
 \left(\partial_{\mu}x^m\partial_\nu x^n \partial_{\rho}x^l C^{[3]}_{mnl}
 +{2\pi\alpha'}{\cal F}_{\mu\nu}
  \partial_\rho x^m C^{[1]}_m
  \right)
~~~.\nn
\eea
The canonical momenta are defined as 
\bea
p_m
&=&-T_{D2}e^{-\phi}\sqrt{-h_F}\left(\frac{1}{2}h_F{}^{(\mu 0)}G_{mn}+
\frac{1}{2}h_F{}^{[\mu 0]}
B_{mn}\right)\partial_\mu x^n\nn\\
&&+T_{D2}\epsilon^{0ij}\left(
\partial_i x^n\partial_j x^l
( \frac{1}{2}C_{mnl}^{[3]}+\frac{1}{3}B_{mn}C^{[1]}_l)
+\frac{1}{3!}2\pi\alpha' {\cal F}_{ij}C_m^{[1]}\right)
\nn\\
E^i&=&-2\pi\alpha' T_{D2}e^{-\phi}
\sqrt{-h_F}\frac{1}{2}h_F{}^{[i0]}+
T_{D2}\frac{1}{3}\epsilon^{0ij}2\pi\alpha'\partial_j x^mC_m^{[1]}
~~~,~~_{i=1,2}~~~.
\eea
The  Hamiltonian and the $\sigma^i$-diffeomorphism constraints are
 given by \cite{Hatsuda:1998by}
\bea
&&\left\{
{\renewcommand{\arraystretch}{1.2}
\begin{array}{ccl}
{\cal H}_\perp &=&\displaystyle\frac{1}{2T_{D2}}e^{\phi}\left(
\tilde{p}_m G^{mn}\tilde{p}_n
+\frac{1}{(2\pi\alpha')^2}
\tilde{E}^ih_{ij}\tilde{E}^j
+T_{D2}^2e^{-2\phi} \det h_F{}_{ij}
\right)~=~0\nn\\
{\cal H}_i&=&\partial_i  x^m\tilde{p}_m+{\cal F}_{ij}\tilde{E}^j~=~
\partial_i  x^m{p}_m+{F}_{ij}{E}^j~=~0\\
\Phi&=&\partial_iE^i
~=~0
\end{array}}\right.
\nn
\eea
with 
\bea
\tilde{p}_m&\equiv&p_m
-B_{mn}\frac{1}{2\pi\alpha'}E^i\partial_i x^n\nn\\
&&-T_{D2}\epsilon^{0ij}\left(
\partial_i x^n\partial_j x^l
( \frac{1}{2}C_{mnl}^{[3]}+\frac{1}{3}B_{mn}C^{[1]}_l)
+\frac{1}{3!}2\pi\alpha' {\cal F}_{ij}C_m^{[1]}\right)
\nn\\
&=&-T_{D2}e^{-\phi}
\sqrt{-h_F}\frac{1}{2}h_F{}^{(\mu 0)}G_{mn}\partial_\mu ^n~~\nn\\
\tilde{E}^i&\equiv&E^i-
T_{D2}\frac{1}{3}\epsilon^{0ij}2\pi\alpha'\partial_j x^mC_m^{[1]}
=-{2\pi\alpha'} T_{D2}e^{-\phi}\sqrt{-h_F}\frac{1}{2} h_F{}^{[i0]}~.
\eea
The determinant term can be rewritten as
\bea
\det h_{Fij}&=&\frac{1}{2}
(\epsilon^{ij}\partial_i x^m \partial_jx^n )
G_{mm'}G_{nn'}
(\epsilon^{i'j'}\partial_{i'} x^{m'} \partial_{j'}x^{n'}) 
+(2\pi\alpha' {\cal F}_{12})^2~~~.
\eea
Therefore the Hamiltonian for D2-brane is given by
\bea
{\cal H}_\perp&=&\frac{1}{2T_{D2}}
Z_M{}^T~{\cal M}^{MN}~Z_N\nn\\
Z_N&=&\left(
{\renewcommand{\arraystretch}{1.2}
\begin{array}{c}
p_n\\
\frac{1}{2\pi \alpha'}E^i\partial_i x^n\\\hline
T_{D2}(2\pi\alpha')\epsilon^{ij}F_{ij}\\
T_{D2}\epsilon^{ij}\partial_i x^n \partial_jx^l
\end{array}}
\right)\nn\\\nn\\
{\cal M}^{MN}&=&({\cal N}^{T})^{M}{}_{L}{\cal M}_0{}^{LK}
{\cal N}_{K}{}^{N}\\
{\cal N}_{K}{}^{N}&=&
\left(
\begin{array}{cc|cc}
\delta_k{}^n&-B_{kn}&-C^{[1]}_k&-C^{[2]}_{knl}\\
0&\delta^k{}_n&0&-C^{[1]}_{[n}\delta^k_{l]}\\
\hline
0&0&1&B_{nl}\\
0&0&0&\delta_n^k \delta^m_l
\end{array}
\right)\nn\\
{\cal M}_0^{LK}&=&
\left(
\begin{array}{cc|cc}
e^{\phi}G^{lk}&0&&\\
0&e^{\phi}G_{lk}&&\\\hline
&&e^{-\phi}&0\\
&&0&e^{-\phi}G_{lk}G_{l'k'}
\end{array}
\right)\nn~~~.
\eea
The $Z_M$ basis is separated into the 
NS-NS sector and the R-R sector by the middle line.
The upper-left part of the ${\cal M}$ matrix 
with ${\cal M}={\cal N}{\cal M}_0{\cal N}$
is the same as the one for the F-string \bref{F1T}.
The parallel direction diffeomorphism constraints ${\cal H}_i$
is rewritten in terms of $Z_M$ basis by contracting with
$ E^i$ and  $ \epsilon^{ij}\partial_jx$
\bea
&&\tilde{E}^i {\cal H}_i~=~\epsilon^{ij}\partial_i x^m {\cal H}_j~=~0\nn\\
&&~~\Rightarrow~~
Z_M\tilde{\rho}^{MN}Z_N=0~~,~~
\tilde{\rho}^{MN}=
\left(\begin{array}{cc|cc}
0&a\delta_n^m&0&b_{[n}\delta^m_{l]}\\
a\delta_m^n&0&-b_m&0\\\hline
0&-b_n&&\\
b_{[m}\delta^n_{l]}&0&&
\end{array}\right)
\eea
where $a,~b_m$ are arbitrary coefficients.

The $Z_M$ algebra is given by
\bea
&
\left\{Z_M(\sigma),Z_N(\sigma')\right\}=i\rho_{MN}^i
\partial_i \delta(\sigma-\sigma')&\nn\\
&\rho_{MN}=\left(
{\renewcommand{\arraystretch}{1.2}
\begin{array}{cc|cc}
0&
\frac{1}{2\pi\alpha'}E^i\delta_m^n&0
&T_{D2}
\epsilon^{ij}\partial_j x^{[n}\delta_m^{l]}\\
\frac{1}{2\pi\alpha'}E^i\delta_n^m&0&
2T_{D2}
\epsilon^{ij}\partial_j x^m
&0\\\hline
0&
2T_{D2}
\epsilon^{ij}\partial_j x^n
&&\\
T_{D2}
\epsilon^{ij}\partial_jx^{[m}\delta_n^{l]}&0&&
\end{array}}
\right)&\label{D2rho}
\eea
where $\rho_{MN}$ is the worldvolume vector.
From the similar analysis in terms of the following notation
 \bea
& 
\hat{\Lambda}=
\lambda+\lambda^\ast+\lambda^{[0]}+\lambda^{[2]}
~\in~
T \oplus T^\ast \oplus \Lambda^{0}T^\ast  \oplus \Lambda^{2}T^\ast 
&\nn\eea
we get  the following extension of 
the Courant bracket to the one for a D2-brane
\bea
[\hat{\Lambda}_1,\hat{\Lambda}_2]_{D2}
&=&[\lambda_1,\lambda_2]+{\cal L}_{\lambda_{[1}}\lambda_{2]}^\ast
+{\cal L}_{\lambda_{[1}}\lambda_{2]}^{[0]}
+{\cal L}_{\lambda_{[1}}\lambda_{2]}^{[2]}
-\frac{1}{2}d\left(\iota_{\lambda_{[1}} \lambda_{2]}^\ast+
\iota_{\lambda_{[1}} \lambda_{2]}^{[0]}
+\iota_{\lambda_{[1}} \lambda_{2]}^{[2]}\right)
\label{D2}\nn
\\
&&+\frac{1}{2}\left(
\lambda^\ast_{[1}\wedge \lambda_{2]}^{[0]}
-\lambda_{[1}^{[0]}d\lambda_{2]}^\ast
\right)\nn
\eea
\bea
&
\left\{
{\renewcommand{\arraystretch}{1.2}
\begin{array}{ccl}
{\cal L}_{\lambda_{1}}\lambda_{2}^{[0]}&=&
\Lambda_{1}^m\partial_m\Lambda_{2}^{[0]} 
T_{D2}(2\pi\alpha')\epsilon^{ij}F_{ij}\\
{\cal L}_{\lambda_{1}}\lambda_{2}^{[2]}&=&
\left(\Lambda_{1}^m\partial_m\Lambda_{2;nl}^{[2]}
 +\partial_{[n|}\Lambda_{1}^{m}\Lambda_{2;m|l]}
\right)T_{D2}\epsilon^{ij}\partial_i
x^n\partial_jx^l\\
d(\iota_{\lambda_{1}}\lambda^{[0]}_{2})&=&
\partial_m(\Lambda_{1}^m\Lambda_{2}^{[0]})~
T_{D2}(2\pi\alpha')\epsilon^{ij}F_{ij}
\\
d(\iota_{\lambda_{1}}\lambda^{[2]}_{2})&=&
\partial_l(\Lambda_{1}^m\Lambda_{2;mn}^{[2]})~
T_{D2}\epsilon^{ij}\partial_ix^n\partial_jx^l\\
\lambda_1^\ast \wedge d\lambda_{2}^{[0]}&=&
\Lambda_{1[n}\partial_{l]}\Lambda^{[0]}_2~T_{D2}\epsilon^{ij}
\partial_ix^n\partial_jx^l
\end{array}}\right.&
\eea
for $K=0$,
and 
\bea
[\hat{\Lambda}_1,\hat{\Lambda}_2]_{D2}
&=&[\lambda_1,\lambda_2]+{\cal L}_{\lambda_{1}}\lambda_{2}^\ast
+{\cal L}_{\lambda_{1}}\lambda_{2}^{[0]}
+{\cal L}_{\lambda_{1}}\lambda_{2}^{[2]}
-\iota_{\lambda_{2}} d\lambda_{1}^\ast-
\iota_{\lambda_{2}} d\lambda_{1}^{[0]}
-\iota_{\lambda_{2}} d\lambda_{1}^{[2]}
\label{D2half}\nn
\\
&&-d\lambda_1^{[0]}\wedge\lambda^\ast_2 +d\lambda_1^\ast\wedge
\lambda_2^{[0]}
\nn
\\
&&\left\{{\renewcommand{\arraystretch}{1.2}
\begin{array}{ccl}
\iota_{\lambda_{2}}d\lambda^{[0]}_{1}&=&
\Lambda_{2}^m\partial_m\Lambda_{1}^{[0]}
T_{D2}(2\pi\alpha')\epsilon^{ij}F_{ij}
\\
\iota_{\lambda_{2}}d\lambda^{[2]}_{1}&=&
\Lambda_{2}^m\partial_{[l|}\Lambda_{1|mn]}^{[2]}
T_{D2}\epsilon^{ij}\partial_ix^n
\partial_jx^l
\end{array}}\right.\label{D2Courant}
\eea
for $K=-1/2$.

The gauge transformation rule is given by the Courant bracket 
in \bref{D2Courant} as
\bea
&{\renewcommand{\arraystretch}{1.2}
\hat{C}^{D2}_a=\left(
\begin{array}{c}e_a{}^m\\e_{ma}\\\hline e_a{}^lC_{l}^{[1]}\\
e_a{}^lC_{mnl}^{[3]}
\end{array}\right)
~~,~~
\hat{\xi}=\left(
\begin{array}{c}\xi^m\\\xi_m\\\hline
{\xi}^{[0]}\\\xi_{mn}^{[2]}
\end{array}\right)}&\nn\\
&\delta_\xi\hat{C}^{D2}_a=\left[
\hat{\xi},\hat{C}^{D2}_a
\right]_{D2}
~\Rightarrow~
\left\{{\renewcommand{\arraystretch}{1.2}
\begin{array}{ccl}
\delta_\xi C^{[1]}&=&{\cal L}_\xi C^{[1]}+d\xi^{[0]}\\
\delta_\xi C^{[3]}&=&{\cal L}_\xi C^{[3]}+C^{[1]}\wedge d\xi+d\xi^{[2]}
+B\wedge d\xi^{[0]}
\end{array}}\right.
\label{D2}~~~
\eea
where $\xi=\xi_m$ is a gauge parameter for $B$. 
\par
\vskip 6mm
\subsection{D$p$-brane}

The action for a D$p$-brane is similar to \bref{D1act}
replacing the WZ term by
\bea
I&=&I_{DBI}+I_{WZ} ~~,~~I_{DBI}=\displaystyle\int_M 
d^p\sigma~{\cal L}_{DBI}
\nn\\
{\cal L}_{DBI}&=&-T_{Dp}e^{-\phi}\sqrt{-h_{F}}~~,~~
h_{F}=\det h_F{}_{\mu\nu}\nn\\
I_{WZ}&=&T_{Dp}\displaystyle\int_M 
 e^{2\pi\alpha'{\cal F}}C^{RR}
\nn\\
h_F{}_{\mu\nu}&=&\partial_\mu x^m \partial_\nu x^nG_{mn}
+2\pi \alpha' {\cal F}_{\mu\nu} \nn\\
F_{\mu\nu}&=&\partial_\mu A_\nu-\partial_\nu A_\mu ~~,~~
{\cal F}_{\mu\nu}~=~F_{\mu\nu}+\frac{1}{2\pi\alpha'}
\partial_\mu x^m\partial_\nu x^n B_{mn}
~~~.\nn
\eea
The canonical momenta are defined as 
\bea
p_m
&=&-T_{Dp}
\sqrt{-h_F}\left(\frac{1}{2}h_F{}^{(\mu 0)}G_{mn}+
\frac{1}{2}h_F{}^{[\mu 0]}
B_{mn}\right)\partial_\mu x^n
+\displaystyle\frac{\partial {\cal L}_{WZ}}{\partial (\partial_0 x^m)}
\nn\\
E^i&=&-2\pi\alpha' T_{Dp}
\sqrt{-h_F}\frac{1}{2}h_F{}^{[i0]}+
\displaystyle\frac{\partial {\cal L}_{WZ}}{\partial F_{0i}}
~~~,~~_{i=1,\cdots,p}~~~.
\eea
The Hamiltonian is given by
\bea
H&=&\displaystyle\int d^p\sigma ~{\cal H}\nn\\
{\cal H}&=&p_m\partial_0 x^m
+E^i\partial_0 A_i-{\cal L}\nn\\
&=&-\frac{1}{\sqrt{-h}h^{00}}{\cal H}_\perp
-\frac{h^{0i}}{h^{00}}{\cal H}_i
-A_0~\Phi\nn\\
&&\left\{{\renewcommand{\arraystretch}{1.2}
\begin{array}{ccl}
{\cal H}_\perp &=&\displaystyle\frac{1}{2T_{Dp}}e^\phi
\left(
\tilde{p}_m G^{mn}\tilde{p}_n
+\frac{1}{(2\pi\alpha')^2}
\tilde{E}^ih_{ij}\tilde{E}^j
+T_{Dp}{}^2e^{-2\phi} \det h_F{}_{ij}
\right)~=~0\nn\\
{\cal H}_i&=&\tilde{p}_m\partial_i  x^m+{\cal F}_{ij}\tilde{E}^j~=~
{p}_m\partial_i  x^m+{F}_{ij}{E}^j~=~
0\\
\Phi&=&\partial_iE^i~=~0
\end{array}}\right.
\nn
\eea
with 
\bea
\tilde{p}_m&\equiv&p_m
-B_{mn}\frac{1}{2\pi\alpha'}E^i\partial_i x^n
-\displaystyle\frac{\partial {\cal L}_{WZ}}{\partial (\partial_0 x^m)}
\nn\\
&=&-T_{Dp}e^{-\phi}\sqrt{-h_F}\frac{1}{2}h_F{}^{(\mu 0)}G_{mn}\partial_\mu x^n~~\nn\\
\tilde{E}^i&\equiv&E^i-
\displaystyle\frac{\partial {\cal L}_{WZ}}{\partial F_{0i}}
=-{2\pi\alpha'} T_{Dp}e^{-\phi}\sqrt{-h_F}\frac{1}{2} h_F{}^{[i0]}~.
\eea
\par\vskip 6mm
\subsubsection{IIA D$p$-brane}
Let us rewrite the Hamiltonian as the sum of bilinears.
For the type IIA theory $p=2q$ is even.
 ${\cal H}_\perp$ for a IIA D$p$-brane
is written by the sum of bilinears \cite{Hatsuda:1998by}
 as
\bea
{\cal H}_\perp&=&\frac{1}{2T_{Dp}}
Z_M{}^T~{\cal M}^{MN}~Z_N\nn
\\
Z_M&=&\left({\renewcommand{\arraystretch}{1.2}
\begin{array}{c}
p_m\\
\frac{1}{2\pi \alpha'}E^i \partial_i x^m
\\\hline
T_{Dp}(2\pi\alpha')^q { F}^q\\
\vdots\\
T_{Dp}(2\pi\alpha')
\epsilon^{i_1\cdots i_p}F_{i_1i_2}\partial_{i_3}x^{m_1}
 \cdots  \partial_{i_p}x^{m_{p-2}}\\
T_{Dp}\epsilon^{i_1\cdots i_p}\partial_{i_1}
x^{m_1} \cdots  \partial_{i_p}x^{m_{p}}
\end{array}}
\right)\nn\\\nn\\
{\cal M}^{MN}&=&({\cal N}^{T})^{M}{}_{L}{\cal M}_0{}^{LK}
{\cal N}_{K}{}^{N}\nn\\
{\cal N}_{K}{}^{N}&=&
\left(
{\renewcommand{\arraystretch}{1.2}
\begin{array}{cc|cccc}
\delta_k^n&-B_{kn}&-C^{[1]}&\cdots&
-\displaystyle\sum_{r=0}^qC^{[p-1-2r]}B^r
&
-\displaystyle\sum_{r=0}^qC^{[p+1-2r]}B^r
\\
0&\delta^k_n&0&\cdots&-\displaystyle\sum_{r=0}^qC^{[p-3-2r]}B^r
&-\displaystyle\sum_{r=0}^qC^{[p-1-2r]}B^r
\\\hline
0&0&1&\cdots&B^{q-1}&B^q\\
&&&\vdots&&\\
0&0&0&\cdots&{\bf 1}&B\\
0&0&0&\cdots&0&{\bf 1}
\end{array}}
\right)\nn\\
{\cal M}_0^{LK}&=&
\left(
\begin{array}{cc|cccc}
e^\phi G^{lk}&&&&&\\
&e^\phi G_{lk}&&&&\\\hline
&&e^{-\phi}&&&\\&&&\cdots&&\\
&&&&e^{-\phi}G_{l_1k_1}\cdots G_{l_{p-2}k_{p-2}}&\\
&&&&&e^{-\phi}G_{l_1k_1}\cdots G_{l_{p}k_{p}}
\end{array}
\right)\nn
\eea
where the NS-NS two-form $B_{mn}=B^{[2]}$ is denoted by $B$.

The worldvolume diffeomorphism constraints ${\cal H}_i=0$
is written in a bilinear form by contracting with 
$\tilde{E}^i$, $\cdots$, $\epsilon^{i_1\cdots i_p}F_{i_1i_2}\cdots
\partial_{i_p}x$,
\bea
&&\tilde{E}^i {\cal H}_{i}=
\epsilon^{i_1\cdots i_{p-1}i}F_{i_1i_2}\cdots
\partial_{i_{p-1}}x{\cal H}_i
=\epsilon^{i_1\cdots i_{p-1}i}\partial_{i_1}x
\cdots
\partial_{i_{p-1}}x
{\cal H}_i
=0\nn\\
&&~~\Rightarrow~Z_M\tilde{\rho}^{MN}Z_N=0\nn\\
&&\tilde{\rho}^{MN}=
\left(\begin{array}{cc|ccccc}
0&a\delta_n^m&0&b_{[2]}^m&\cdots&\cdots&
c_{[p]}^m\\
a\delta_m^n&0&
b_{[0]m}
&\cdots&\cdots&
c_{[p-2]m}
&0\\\hline
0&b_{[0]n}&&&&&\\
b_{[2]}^n&\cdot&&&&&\\
\vdots&\vdots&&&&&\\
\cdot&c_{[p-2]n}&&&&&\\
c_{[p]}^n&0&&&&
\end{array}\right)\nn\\
&&~~~~~b_{[2]}^m=\beta_{[n}\delta_{l]}^m~,~~~~~~~~~~~~~~~~~~
b_{[0]m}=-\beta_m\nn\\
&&~~~~~c_{[p]}^m=\gamma_{[n_1\cdots n_{p-1}}\delta_{n_p]}^m~,~~~
~~~~~~c_{[p-2]m}=-_pC_2~\gamma_{[n_1\cdots n_{p-2} m]}\nn
\eea
with arbitrary coefficients $a,~\beta,~\gamma$.

The $Z_M$ algebra is given by
\bea
&&
\left\{Z_M(\sigma),Z_N(\sigma')\right\}=
i\rho_{MN}^i\partial_i \delta^{(p)}(\sigma-\sigma')\nn\\\nn\\
&&\rho_{MN}^i=\left({\renewcommand{\arraystretch}{1.2}
\begin{array}{cc|ccccc}
0&
\frac{1}{2\pi\alpha'}E^i\delta_m^n&0
&\rho_{14}&\cdots&&\rho_{16}\\
\frac{1}{2\pi\alpha'}E^i\delta_n^m&0&\rho_{23}&\cdots&&\rho_{25}&0\\
\hline 0&\rho_{23}&&&&&\\
\rho_{14}
&\vdots&&&&&\\
\vdots
&\rho_{25}&&&&&\\
\rho_{16}
&0&&&&\\
\end{array}}
\right)\label{D$p$rho}\nn\\
&&~~~~~~~\rho_{14}=T_{Dp}\epsilon^{ii_1\cdots i_{p-1}}(2\pi\alpha')^{q-1}
F^{q-1}_{i_1\cdots i_{p-2}}\partial_{i_{p-1}}x^{[n_1}\delta_m^{n_{2}]}\nn\\
&&~~~~~~~\rho_{23}=T_{Dp}\epsilon^{ii_1\cdots i_{p-1}}(2\pi\alpha')^{q-1}
F^{q-1}_{i_1\cdots i_{p-2}}\partial_{i_{p-1}}x^m\nn\\
&&~~~~~~~\rho_{16}=T_{Dp} 
\epsilon^{ii_1\cdots i_{p-1}}\partial_{i_{1}}
x^{[n_1}\cdots \partial_{i_{p-1}}x^{n_{p-1}}\delta_m^{n_{p}]}
\nn\\
&&~~~~~~~\rho_{25}=T_{Dp} 
\epsilon^{ii_1\cdots i_{p-1}}\partial_{i_{1}}
x^{[n_1}\cdots \partial_{i_{p-1}}x^{m]}
\nn
\eea
where $\rho_{MN}^i$ is a worldvolume vector.
From the similar analysis we get  the following extension of 
the Courant bracket to the one for a IIA D$p$-brane
\bea
& 
\hat{\Lambda}=
\lambda+\lambda^\ast+\lambda^{[0]}+\cdots+\lambda^{[p]}
~\in~
T \oplus T^\ast \oplus \Lambda^{0}T^\ast  \oplus\cdots \oplus
 \Lambda^{p}T^\ast 
&\nn
\eea
\bea
\left[\hat{\Lambda}_1,\hat{\Lambda}_2\right]_{Dp}&=&
[\lambda_1,\lambda_2]+{\cal L}_{\lambda_{[1}}\lambda_{2]}^\ast
+{\cal L}_{\lambda_{[1}}\lambda_{2]}^{[0]}
+\cdots+
{\cal L}_{\lambda_{[1}}\lambda_{2]}^{[p]}
\nn\\
&&-\frac{1}{2}d(\iota_{\lambda_{[1}} \lambda_{2]}^\ast+
\iota_{\lambda_{[1}} \lambda_{2]}^{[0]}
+\cdots+\iota_{\lambda_{[1}} \lambda_{2]}^{[p]})\nn\\
&&
+\displaystyle\sum_{s=1}^{q}
\frac{s}{(p+2-2s)!}
(\lambda^\ast_{[1}\wedge d\lambda_{2]}^{[p-2s]}
+d\lambda_{[1}^\ast \wedge \lambda_{2]}^{[p-2s]}
)
~~~{\rm for}~K=0\label{IIADp}\\
\left[\hat{\Lambda}_1,\hat{\Lambda}_2\right]_{Dp}
&=&[\lambda_1,\lambda_2]+{\cal L}_{\lambda_{1}}\lambda_{2}^\ast
+{\cal L}_{\lambda_{1}}\lambda_{2}^{[0]}
+\cdots+{\cal L}_{\lambda_{1}}\lambda_{2}^{[p]}
\nn\\
&&-\iota_{\lambda_{2}} d\lambda_{1}^\ast-
\iota_{\lambda_{2}} d\lambda_{1}^{[0]}
-\cdots -\iota_{\lambda_{2}} d\lambda_{1}^{[p]}\nn\\
&&+\displaystyle\sum_{s=1}^{q}
\frac{2s}{(p+2-2s)!}
(d\lambda^\ast_{1}\wedge d\lambda_{2}^{[p-2s]}
-d\lambda_{1}^{[p-2s]} \wedge \lambda_{2}^\ast
)
~~~{\rm for}~K=-\frac{1}{2}
\label{IIAD$p$half}
\eea
The gauge transformation rule for the R-R gauge fields 
is given by the Courant bracket 
in \bref{IIAD$p$half} as
\bea
&{\renewcommand{\arraystretch}{1.2}
\hat{C}^{Dp}_a=\left(
\begin{array}{c}e_a{}^m\\e_{ma}\\\hline
e_a{}^lC_{l}^{[1]}\\\vdots\\
e_a{}^lC_{m_1\cdots m_p l}^{[p+1]}
\end{array}\right)
~~,~~
\hat{\xi}=\left(
\begin{array}{c}\xi^m\\\xi_m\\\hline
{\xi}^{[0]}\\
\vdots\\
\xi_{m_1\cdots m_p}^{[p]}
\end{array}\right)}&\nn\\
\nn\\
&\delta_\xi\hat{C}^{Dp}_a=\left[
\hat{\xi},\hat{C}^{Dp}_a
\right]_{Dp}
~\Rightarrow~
\left\{\begin{array}{ccl}
\delta_\xi C^{[1]}&=&{\cal L}_\xi C^{[1]}+d\xi^{[0]}\\
\vdots\\
\delta_\xi C^{[p+1]}&=&{\cal L}_\xi C^{[p+1]}
+C^{[p-1]}\wedge d\xi
+d\xi^{[p]}+B\wedge d\xi^{[p-2]}
\end{array}\right.
\label{IIADpbrane}~~~
\eea
where $\xi=\xi_m$ is a gauge parameter for $B$.

\par
\vskip 6mm
\subsubsection{IIB D$p$-brane}

For the type IIB $p=2q+1$ is odd.
Then   ${\cal H}_\perp$ for a IIB D$p$-brane
is written by the sum of bilinears
\cite{Kamimura:1997ju}
 as
\bea
{\cal H}_\perp&=&\frac{1}{2T_{Dp}}
Z_M{}^T~{\cal M}^{MN}~Z_N\nn\\
Z_M&=&\left({\renewcommand{\arraystretch}{1.2}
\begin{array}{c}
p_m\\
\frac{1}{2\pi \alpha'}E^i\partial_ix^m
\\\hline
T_{Dp}(2\pi\alpha')^q { F}^q \wedge dx^m\\
\vdots\\
T_{Dp}(2\pi\alpha')
\epsilon^{i_1\cdots i_p} { F}_{i_1i_2}
\partial_{i_3}x^{m_1}\cdots \partial_{i_p}x^{m_{p-2}}\\
T_{Dp}
\epsilon^{i_1\cdots i_p}\partial_{i_1}x^{m_1}
\cdots \partial_{i_p}x^{m_p}
\end{array}}
\right)\nn\\
\nn\\
{\cal M}^{MN}&=&({\cal N}^{T})^{M}{}_{L}{\cal M}_0{}^{LK}
{\cal N}_{K}{}^{N}\nn \\
{\cal N}_{K}{}^{N}&=& \nn
\left({\renewcommand{\arraystretch}{1.2}
\begin{array}{cc|cccc}
\delta_k{}^n&-B_{kn}&-C^{[2]}-C^{[0]}B&\cdots&
-\displaystyle\sum_{r=0}^qC^{[p-1-2r]}B^r
&
-\displaystyle\sum_{r=0}^qC^{[p+1-2r]}B^r
\\
0&\delta^k{}_n&C^{[0]}&\cdots&
-\displaystyle\sum_{r=0}^qC^{[p-3-2r]}B^r
&-\displaystyle\sum_{r=0}^qC^{[p-1-2r]}B^r
\\\hline
0&0&1&\cdots&B^{q-1}&B^q\\
&&&\vdots&&\\
0&0&0&\cdots&{\bf 1}&B\\
0&0&0&\cdots&0&{\bf 1}
\end{array}}
\right)\\
{\cal M}_0^{LK}&=&
\left(
\begin{array}{cc|cccc}
e^\phi G^{lk}&&&&&\\
&e^\phi G_{lk}&&&&\\\hline
&&e^{-\phi}G_{lk}&&&\\&&&\cdots&&\\
&&&&e^{-\phi}G_{l_1k_1}\cdots G_{l_{p-2}k_{p-2}}&\\
&&&&&e^{-\phi}G_{l_1k_1}\cdots G_{l_{p}k_{p}}
\end{array}
\right)\nn\\
\label{IIBM0}
\eea

The worldvolume diffeomorphism constraints ${\cal H}_i=0$
is written in a bilinear form by contarcting with 
$E^i$, $\cdots$, 
$\epsilon^{i_1\cdots i_p}F_{i_1i_2}\cdots \partial_{i_p}x$,
\bea
&&\tilde{E}^i {\cal H}_{i}=
\epsilon^{ii_1\cdots i_{p-1}}\partial_{i_1} x
\cdots
\partial_{i_{p-1}}x
{\cal H}_{i}
=
\epsilon^{ii_1\cdots i_{p-1}}
F_{i_1i_2}\cdots
\partial_{i_{p-1}}x {\cal H}_i=0\nn\\
&&~~
Z_M\tilde{\rho}^{MN}Z_N=0~~\nn\\~~&&
\tilde{\rho}^{MN}=
\left(\begin{array}{cc|ccc}
0&a\delta_n^m&b\delta^m_{n}&\cdots&
c_{[p]}^m\\
a\delta_m^n&0&\cdots&c_{[p-2]m}
&0\\\hline
b\delta^n_{m}&\vdots&&&\\
\vdots&c_{[p-2]n}&&&\\
c_{[p]}^n&0&&&
\end{array}\right)\nn\\
&&~~~~~c_{[p]}^m=\gamma_{[n_1\cdots n_{p-1}}\delta_{n_p]}^m~,~~~
~~~~~~c_{[p-2]m}=-_pC_2~\gamma_{[n_1\cdots n_{p-2} m]}\nn
\eea
with arbitrary coefficients $a,~b,~\gamma$.

The $Z_M$ algebra is given by
\bea
&&
\left\{Z_M(\sigma),Z_N(\sigma')\right\}=
i\rho_{MN}^i \partial_i\delta^{(p)}(\sigma-\sigma')\nn\\\nn\\
&&\rho_{MN}^i=\left({\renewcommand{\arraystretch}{1.2}
\begin{array}{cc|cccc}
0&
\frac{1}{2\pi\alpha'}E^i\delta_m^n
&\rho_{13}&\cdots&&\rho_{15}\\
\frac{1}{2\pi\alpha'}E^i\delta_n^m&0&\cdots&&\rho_{24}&0\\\hline
\rho_{13}
&\vdots&&&\\
\vdots&\rho_{24}&&&\\
\rho_{15}
&0&&&\\
\end{array}}
\right)\label{IIBD$p$rho}\nn\\
&&~~~~~~~
\rho_{13}=T_{Dp}\epsilon^{ii_1\cdots i_{p-1}}(2\pi\alpha')^{q}(F^{q}){}_{i_1\cdots i_{p-2}} 
\delta_m^{n}\nn\\
&&~~~~~~~\rho_{15}=T_{Dp} \epsilon^{ii_1\cdots i_{p-1}}
\partial_{i_1}x^{[n_1}\cdots \partial_{i_{p-1}}x^{n_{p-1}}\delta_m^{n_{p}]}\nn\\
&&~~~~~~~\rho_{24}=T_{Dp} \epsilon^{ii_1\cdots i_{p-1}}
\partial_{i_1}x^{[m}\cdots \partial_{i_{p-1}}x^{n_{p-2}]}
\nn
\eea
where $\rho_{MN}^i$ is a worldvolume vector.
The Courant bracket to the one for a IIB D$p$-brane
in terms of a vector as
\bea
& 
\hat{\Lambda}=
\lambda+\lambda^\ast+\lambda^{[1]}+\cdots+\lambda^{[p]}
~\in~
T \oplus T^\ast \oplus \Lambda^{1}T^\ast  \oplus\cdots \oplus
 \Lambda^{p}T^\ast 
&\nn
\eea
\bea
\left[\hat{\Lambda}_1,\hat{\Lambda}_2\right]_{Dp}&=&
[\lambda_1,\lambda_2]+{\cal L}_{\lambda_{[1}}\lambda_{2]}^\ast
+{\cal L}_{\lambda_{[1}}\lambda_{2]}^{[1]}
+\cdots+
{\cal L}_{\lambda_{[1}}\lambda_{2]}^{[p]}
\nn\\
&&-\frac{1}{2}d(\iota_{\lambda_{[1}} \lambda_{2]}^\ast+
\iota_{\lambda_{[1}} \lambda_{2]}^{[1]}
+\cdots+\iota_{\lambda_{[1}} \lambda_{2]}^{[p]})
\nn\\&&+
\displaystyle\sum_{s=1}^{q}
\frac{s}{(p+2-2s)!}
(\lambda^\ast_{[1}\wedge d\lambda_{2]}^{[p-2s]}
+d\lambda_{[1}^\ast \wedge \lambda_{2]}^{[p-2s]}
)
~~{\rm for}~K=0\label{IIBDp}\\
\left[\hat{\Lambda}_1,\hat{\Lambda}_2\right]_{Dp}
&=&[\lambda_1,\lambda_2]+{\cal L}_{\lambda_{1}}\lambda_{2}^\ast
+{\cal L}_{\lambda_{1}}\lambda_{2}^{[1]}
+\cdots+{\cal L}_{\lambda_{1}}\lambda_{2}^{[p]}
\nn\\
&&-\iota_{\lambda_{2}} d\lambda_{1}^\ast-
\iota_{\lambda_{2}} d\lambda_{1}^{[1]}
-\cdots -\iota_{\lambda_{2}} d\lambda_{1}^{[p]}
\nn\\
&&+
\displaystyle\sum_{s=1}^{q}
\frac{2s}{(p+2-2s)!}
(d\lambda^\ast_{1}\wedge \lambda_{2}^{[p-2s]}
-d\lambda_{1}^{[p-2s]} \wedge \lambda_{2}^\ast
)
~~{\rm for}~K=-\frac{1}{2}
\label{IIBDphalf}
\eea
The gauge transformation rule for the R-R gauge fields 
is given by the Courant bracket 
in \bref{IIBDphalf} as
\bea
&{\renewcommand{\arraystretch}{1.2}
\hat{C}^{Dp}_a=\left(
\begin{array}{c}e_a{}^m\\e_{ma}\\\hline
e_a{}^lC_{ml}^{[2]}\\\vdots\\
e_a{}^lC_{m_1\cdots m_p l}^{[p+1]}
\end{array}\right)
~~,~~
\hat{\xi}=\left(
\begin{array}{c}\xi^m\\\xi_m\\\hline
{\xi}^{[1]}_m\\
\vdots\\
\xi_{m_1\cdots m_p}^{[p]}
\end{array}\right)}&\nn\\
\nn\\
&\delta_\xi\hat{C}^{Dp}_a=\left[
\hat{\xi},\hat{C}^{Dp}_a
\right]_{Dp}
~\Rightarrow~
\left\{\begin{array}{ccl}
\delta_\xi C^{[2]}&=&{\cal L}_\xi C^{[2]}+d\xi^{[1]}\\
\vdots\\
\delta_\xi C^{[p+1]}&=&{\cal L}_\xi C^{[p+1]}
-C^{[p-1]}\wedge d\xi+d\xi^{[p]}+B\wedge d\xi^{[p-2]}
\end{array}\right.
\label{IIBDpbrane}~~~
\eea
where $\xi=\xi_m$ is a gauge parameter for $B$.

\par
\vskip 6mm
\section*{Acknowledgements}
We are grateful to the JHEP referee for the critical advice.
We are also grateful to Kiyoshi Kamimura for helpful discussions
especially on the canonical treatment of the DBI fields.
We would like to thank Warren Siegel for fruitful discussions 
and introducing references. 
The work of M.H. is supported  by Grant-in-Aid for Scientific Research (C) No. 24540284 from The Ministry of Education, Culture, Sports, Science and Technology of Japan.
\vskip 6mm
\appendix
\section*{Appendix}
\par
\vskip 6mm

\section{Determinant}

In this appendix we show that the  det$h_{Fij}$ 
is written in a bilinear form,
where $h_{Fij}=h_{ij}+f_{ij}$ 
 contains a symmetric matrix $h_{ij}=\partial_ix^m\partial_jx^nG_{mn}$ and an antisymmetric matrix $f_{ij}=2\pi\alpha'F_{ij}+\partial_ix^m\partial_jx^nB_{mn}$.
The coefficients of each term are determined.

 As a nontrivial example  we begin by a determinant 
 det$h_{Fij}$ for $p=4$ case  with
 $i,j=1,\cdots,4$;
 \bea
 \det h_{Fij}&=&
 \frac{1}{4!}\epsilon^{ijkl} h_{ii'}h_{jj'}h_{kk'}h_{ll'}\epsilon^{i'j'k'l'}\nn\\
&&+\frac{_4C_2}{4!}\epsilon^{ijkl} h_{ii'}h_{jj'}f_{kk'}f_{ll'}\epsilon^{i'j'k'l'}\nn\\
&&+\frac{1}{4!}\epsilon^{ijkl} f_{ii'}f_{jj'}f_{kk'}f_{ll'}\epsilon^{i'j'k'l'}\nn\\
&=&\frac{1}{4!}\epsilon^{ijkl}\partial_ix^{m_1}\partial_ix^{m_2}
\partial_ix^{m_3}\partial_ix^{m_4}
G_{m_1n_1}G_{m_2n_2}G_{m_3n_3}G_{m_4n_4}
\epsilon^{i'j'k'l'}\partial_{i'}x^{n_1}
\partial_{j'}x^{n_2}\partial_{k'}x^{n_3}\partial_{l'}x^{n_4}\nn\\
&&+\frac{_4C_2}{4!\cdot 2}
\epsilon^{ijkl}\partial_ix^{m_1}\partial_jx^{m_2}f_{kl}
G_{m_1n_1}G_{m_2n_2}
\epsilon^{i'j'k'l'}\partial_{i'}x^{n_1}
\partial_{j'}x^{n_2} f_{k'l'}\nn\\
&&+\frac{3}{4!\cdot 2\cdot 4}
\epsilon^{ijkl}f_{ij}f_{kl}
\epsilon^{i'j'k'l'}\partial_{i'}x^{n_1}
f_{i'j'} f_{k'l'}\nn~~~.
 \eea
Useful relation
obtained by taking totally antisymmetric indices for five indices
is the following: 
\bea
0&=&\epsilon^{ijkl}\epsilon^{i'j'k'l'}
\left(h_{ii'}h_{jj'}f_{kk'}f_{ll'}
+4{\rm terms~totally~antisymmetric~in}~{ijkll'}\right)
\nn\\
&=&\epsilon^{ijkl}\epsilon^{i'j'k'l'}
\left(2h_{ii'}h_{jj'}f_{kk'}f_{ll'}
-h_{kk'}h_{ll'}f_{i'j'}f_{ij}\right)\nn\\
\Rightarrow
&&\epsilon^{ijkl}\epsilon^{i'j'k'l'}
h_{ii'}h_{jj'}f_{kk'}f_{ll'}\nn\\
&=&\displaystyle\frac{1}{2}
\left(\epsilon^{ijkl}\partial_{i}x^{m_1}
\partial_{j}x^{m_2}f_{kl}\right)G_{m_1n_1}G_{m_2n_2}
\left(\epsilon^{i'j'k'l'}\partial_{i'}x^{n_1}
\partial_{j'}x^{n_2}f_{k'l'}\right)~~~.
\eea
Analogously the factorization of each term 
can be confirmed.
 
Systematic analysis of generic dimension is given as follows.
The generic form of a D$p$-brane in type IIA: ($p = 2q$)
\bea
\begin{array}{cl}
\det h_{Fij} \ &= \ 
\frac{1}{p!} 
G_{m_1 n_1} \cdots G_{m_p n_p} 
\Big( \eps^{i_1 \cdots i_p} \del_{i_1} x^{m_1} \cdots \del_{i_p} x^{m_p} \Big)^2
  \\
\ & \ \ \ \ 
+ \frac{1}{(2^1)^2} 
G_{m_1 m'_1} \cdots G_{m_{p-2} m'_{p-2}}
\Big( \eps^{i_1 \cdots i_p} \del_{i_1} x^{m_1} \cdots \del_{i_{p-2}} x^{m_{p-2}} f_{i_{p-1} i_p} \Big)^2
  \\
\ & \ \ \ \ 
+ \frac{1}{(2^3)^2} 
G_{m_1 m'_1} \cdots G_{m_{p-4} m'_{p-4}}
\Big( \eps^{i_1 \cdots i_p} \del_{i_1} x^{m_1} \cdots \del_{i_{p-4}} x^{m_{p-4}} f_{i_{p-3} i_{p-2}} f_{i_{p-1} i_p} \Big)^2
  \\
\ & \ \ \ \ 
+ \ldots
  \\
\ & \ \ \ \ 
+ \frac{1}{(2^{\alpha_q})^2} 
\Big( \eps^{i_1 \cdots i_p} f_{i_1 i_2} \cdots f_{i_{p-1} i_p} \Big)^2
  \\
\ &= \ 
\frac{1}{p!} 
G_{m_1 n_1} \cdots G_{m_p n_p} 
\Big( \eps^{i_1 \cdots i_p} \del_{i_1} x^{m_1} \cdots \del_{i_p} x^{m_p} \Big)^2
  \\
\ & \ \ \ \ 
+ \sum_{k=1}^q \frac{1}{(2^{\alpha_k})^2}
G_{m_{2k+1} m'_{2k+1}} \cdots G_{m_{p} m'_{p}}
\Big[ \eps^{i_1 \cdots i_p} 
\big( f_{i_1 i_{2}} \cdots f_{i_{2k-1} i_{2k}} \big) 
\del_{i_{2k+1}} x^{m_{2k+1}} \cdots \del_{i_p} x^{m_p} 
\Big]^2
\, , \\
\alpha_k \ &= \ 
_{2k} C_2 \times \, _{2k-2} C_2 \times \cdots \times \, _2 C_2 \times \frac{1}{k!}
\ = \ 
\frac{(2k)!}{(2!)^k k!}
\, , \\
\alpha_q \ &= \ 
_{2q} C_2 \times \, _{2q-2} C_2 \times \cdots \times \, _2 C_2 \times \frac{1}{q!}
\ = \ 
\frac{(2q)!}{(2!)^q q!}
\, .
\end{array}\label{coefficient1}
\eea
\par
\vskip 6mm
\noindent
The generic form of a D$p$-brane in type IIB: ($p = 2q + 1$)
\bea
\begin{array}{cl}
\det h_{Fij} \ &= \ 
\frac{1}{p!} 
G_{m_1 n_1} \cdots G_{m_p n_p} 
\Big( \eps^{i_1 \cdots i_p} \del_{i_1} x^{m_1} \cdots \del_{i_p} x^{m_p} \Big)^2
  \\
\ & \ \ \ \ 
+ \frac{1}{(2^1)^2} 
G_{m_1 m'_1} \cdots G_{m_{p-2} m'_{p-2}}
\Big( \eps^{i_1 \cdots i_p} \del_{i_1} x^{m_1} \cdots \del_{i_{p-2}} x^{m_{p-2}} f_{i_{p-1} i_p} \Big)^2
  \\
\ & \ \ \ \ 
+ \frac{1}{(2^3)^2} 
G_{m_1 m'_1} \cdots G_{m_{p-4} m'_{p-4}}
\Big( \eps^{i_1 \cdots i_p} \del_{i_1} x^{m_1} \cdots \del_{i_{p-4}} x^{m_{p-4}} f_{i_{p-3} i_{p-2}} f_{i_{p-1} i_p} \Big)^2
  \\
\ & \ \ \ \ 
+ \ldots
  \\
\ & \ \ \ \ 
+ \frac{1}{(2^{\alpha_q})^2} G_{mm'}
\Big( \eps^{i_1 \cdots i_p} f_{i_1 i_2} \cdots f_{i_{2q-1} i_{2q}} \del_{i_{2q+1}} x^m \Big)^2
  \\
\ &= \ 
\frac{1}{p!} 
G_{m_1 n_1} \cdots G_{m_p n_p} 
\Big( \eps^{i_1 \cdots i_p} \del_{i_1} x^{m_1} \cdots \del_{i_p} x^{m_p} \Big)^2
  \\
\ & \ \ \ \ 
+ \sum_{k=1}^q \frac{1}{(2^{\alpha_k})^2}
G_{m_{2k+1} m'_{2k+1}} \cdots G_{m_{p} m'_{p}}
\Big[ \eps^{i_1 \cdots i_p} 
\big( f_{i_1 i_{2}} \cdots f_{i_{2k-1} i_{2k}} \big) 
\del_{i_{2k+1}} x^{m_{2k+1}} \cdots \del_{i_p} x^{m_p} 
\Big]^2
\, , \\
\alpha_k \ &= \ 
_{2k} C_2 \times \, _{2k-2} C_2 \times \cdots \times \, _2 C_2 \times \frac{1}{k!}
\ = \ 
\frac{(2k)!}{(2!)^k k!}
\, .
\end{array}
\label{coefficient2}
\eea
Formally the determinant in type IIB has the same representation as to the one in type IIA.

\par

\end{document}